\documentclass[12pt]{article}
\usepackage{amssymb}
\usepackage{amsmath}
\usepackage{graphicx}

\begin{document}

\begin{center}
\textbf{Stability and instability of Ellis and phantom wormholes: Are there ghosts?}

\bigskip

K.K. Nandi$^{1,2,3,a}$, A.A. Potapov$^{3,b}$, R.N. Izmailov$^{2,c}$, A.Tamang%
$^{4,d}$

and

J.C. Evans$^{5,e}$

\bigskip

$^{1}$Department of Mathematics, University of North Bengal, Siliguri
734013, WB, India \\[0pt]

$^{2}$Zel'dovich International Center for Astrophysics, M. Akmullah Bashkir
State Pedagogical University, Ufa 450000, RB, Russia

$^{3}$Department of Physics \& Astronomy, Bashkir State University,
Sterlitamak Campus, Sterlitamak 453103, RB, Russia

$^{4}$Department of Mathematics, Darjeeling Government College, Richmond
Hill, Darjeeling 734104, WB, India\\[0pt]
$^{5}$Program in Science, Technology and Society, University of Puget Sound,
1500 North Warner Street,

Tacoma, WA 98416, USA

\bigskip

\bigskip

$^{a}$E-mail: kamalnandi1952@yahoo.co.in

$^{b}$E-mail: potapovaa@mail.ru

$^{c}$E-mail: izmailov.ramil@gmail.com

$^{d}$E-mail: amarjit.tamang1986@gmail.com

$^{e}$E-mail: jcevans@pugetsound.edu

$\bigskip $

-------------------------------------------------------------------------------------------

\bigskip

\textbf{Abstract}
\end{center}

It is concluded in the literature that Ellis wormhole is unstable under
small perturbations and would decay either to the Schwarzschild black hole
or expand away to infinity. While this deterministic conclusion of
instability is correct, we show that the Ellis wormhole reduces to
Schwarzschild black hole \textit{only} when the Ellis solution parameter 
$\gamma $ assumes a complex value $-i$. We shall then reexamine stability of
Ellis and phantom wormholes from the viewpoint of local and asymptotic
observers by using a completely different approach, viz., we adapt
Tangherlini's nondeterministic, prequantal statistical simulation about
photon motion in the real optical medium to an effective medium
reformulation of motions obtained via Hamilton's optical-mechanical analogy
in a gravity field. A crucial component of Tangherlini's idea is the
observed increase of momentum of the photons entering a real medium. We show
that this fact has a heuristic parallel in the effective medium\ version of
the Pound-Rebka experiment in gravity. Our conclusion is that there is a
non-zero probability that Ellis and phantom wormholes could appear stable or
unstable depending on the location of observers and on the values of 
$\gamma$, leading to the possibility of \textit{ghost wormholes} (like ghost
stars). The Schwarzschild horizon, however, would always appear certainly stable
($R=1$, $T=0$) to observers regardless of their location. Phantom
wormholes of bounded mass in the extreme limit $a\rightarrow -1$ are also
shown to be stable just as the Schwarzschild black hole is. We shall propose
a thought experiment showing that our non-deterministic results could be
numerically translated into observable deterministic signatures of ghost
wormholes.

\bigskip


\bigskip

\bigskip

\textbf{1. Introduction}

About forty years ago, Tangherlini [1] using heuristic arguments derived the
Fresnel reflectivity ($R$) and transmissivity ($T$) coefficients for a
photon entering from vacuum into a homogeneous, isotropic, semi-infinite
real medium. The method depends in part on the behavior of the momentum of
light in ordinary refractive media, and in part on the assumption of a
probabilistic condition at the surface where the photon impinges. And why
today, after so many years, this work is recalled at all? The reason is that
Tangherlini's approach could provide a new method of studying an issue of
current interest, namely, the stability of static spherically symmetric
solutions of general relativity. It would be of interest to explore the
impact of Tangherlini's idea of "non-classical, prequantal statistical
simulation" of reflectivity ($R$) and transmissivity ($T$) coefficients on
the (in)stability of the Ellis [2] and phantom wormhole [3] solutions.

The static, spherically symmetric Ellis wormhole, independently derived also
by Bronnikov [4], is an excellent example of a natural (as opposed to
artificially assembled), everywhere regular, traversable wormhole solution
of Einstein's equations sourced by a massless ghost scalar field that has a
negative sign before the kinetic term\footnote{%
The solution should be appropriately termed as the Ellis-Bronnikov wormhole
but since it is more commonly known as Ellis wormhole, such as in Ref.[6],
we keep to this common nomenclature in order to avoid confusion.}. Gonz\'{a}%
lez, Guzm\'{a}n and Sarbach [5] have shown that both the massless and
massive wormholes are linearly unstable for the more general class of
spherically symmetric perturbations, which do not necessarily vanish at the
throat. The authors, in a subsequent paper [6], showed that Ellis wormholes
are unstable also under non-linear perturbations such that the wormhole
either expands away rapidly or collapses to a Schwarzschild black hole. The
stability analysis has been extended by Bronnikov, Fabris and Zhidenko [7]
to small radial perturbations of scalar-vacuum configurations with arbitrary
external potential $V(\phi )$. They concluded the same instability. See also
[8].

There is also an exception in the literature: Previously, Armend\'{a}riz-Pic%
\'{o}n [9] showed that massless Ellis wormhole and at least a non-zero
measure set of massive Ellis wormholes are stable. But it is subsequently
argued by Gonz\'{a}lez, Guzm\'{a}n and Sarbach [5] that the linear stability
analysis in [9] is "incomplete in the sense that it considers only a
restricted class of perturbations where the areal radius of the wormhole
throat is unperturbed and the perturbed scalar field vanishes on the throat.
Such perturbations are somewhat artificial because one could imagine
perturbing the scalar field by a small ingoing pulse which is supported away
from the throat at some time $t=0$, say. As $t$ grows, this pulse travels
towards the throat and since the wormhole metric is everywhere regular one
expects it to reach and cross the throat at some finite time. On the other
hand, requiring that the perturbed scalar field ($\delta \phi $) vanishes at
the throat corresponds to \textit{placing a mirror at the throat which
reflects the scalar pulse.}" We shall take this quote as a guiding line and
(in)stability of a wormhole will be determined by the probability of
reflection of pulses by the wormhole throat.

The analysis in [5] uses perturbation of the scalar field by a small ingoing
pulse of unidentified physical origin because from the mathematical
viewpoint such identification is not mandatory. However, from the physical
viewpoint, two issues could be raised: what causes the perturbation and who
observes the resultant instability of the Ellis wormhole? To that end, we
shall assume that the perturbation $\delta \phi $ of the scalar field $\phi $
is realistically caused by the motion of photons and matter particles.
Further, we shall pay due consideration to the location of observers. In
general relativity, both the issues are of paramount importance because
motion of objects in the gravity field depends on their asymptotic
velocities $-$ for photons it is always unchangeably $c_{0}$, whereas for
particles it can always be changed at will to any value less than $c_{0}$.
Likewise, observations depend on the location that determines the scales and
clocks of the observer for the measurement of an observable there. Since the
master equation in [5] is expressed in the "coordinate" ($t,x,\theta
,\varphi $) language, the instability then is supposed to be observed by
asymptotic observers, whose scales and clocks are unaffected by gravity%
\footnote{%
Two types of quantities should be distinguished in a gravity field, hence by
default in the effective optical medium (see Sec.3), which are: (i)
"Coordinate" quantities (denoted here by primes and observed by asymptotic
observers) are defined by scales and clocks unaffected by gravity, e.g., $%
c^{\prime }(\mathbf{r)=}\left\vert \frac{d\mathbf{r}}{dt}\right\vert =\frac{%
c_{0}}{n(\mathbf{r})}$ [see Eq.(8)]. This is nothing but the \textit{exact }%
equation [10] for Shapiro time delay\ ($\Delta t$) measured by observers
that are physically located at almost asymptotic locations, where proper
scales are nearly equal to coordinate scales. Also, the familiar geodesic
equations for planetary precession, light deflection etc are all expressed
in coordinate language with the effects measured by observers stationed at
asymptotic locations. (ii) "Proper" quantities (denoted here by tilde and
observed by local observers) are defined by scales and clocks affected by
gravity, e.g., the proper length $dL=\Phi ^{-1}\left\vert d\mathbf{r}%
\right\vert $ and proper time $d\tau =\Omega dt$, so that $\widetilde{c}(%
\mathbf{r})=\left\vert \frac{dL}{d\tau }\right\vert =c_{0}$. (iii) In the
absence of gravity or medium \textit{everywhere}, we have what we call here
"free space" quantities denoted without primes or tildes ($n=1$) for which $%
c(\mathbf{r)=}\left\vert \frac{d\mathbf{r}}{dt}\right\vert =c_{0}$.}.

The motivation of and the methodology for this work are as follows. Gravity
is perceived as an \textit{effective refractive medium}\footnote{%
The effective refractive medium approach to gravity for light propagation
was first pointed out by Eddington [11], an idea further developed by de
Felice [12] and later extended to include both light and massive particle
motions [13-18]. The works include WKB approximation [14], quantum phase
[15], Fresnel drag coefficients [16] and cosmology [17]. Going a step
further, even the rotational Kerr metric could be framed as a rotating
medium with the force on a particle analogous to Lorentz force of
electrodynamics [18].} by different observers observing the photons travel
towards the throat and reflected off or transmitted through it. It will be
seen that the equations valid in the real medium remarkably resemble the
equations derived in the effective medium, and supported by experiment in
the latter. It is exactly this situation that motivates us to employ
Tangherlini's probabilistic coefficients in the effective medium. It is
further shown in [1] that, as a result of a certain cancellation in the
statistical simulation, arising out of the independence of collisions, the
coefficients are the \textit{same} for one photon as for a large number of
photons. This result is of importance for the interpretation of
probabilistic coefficients at the wormhole throat. We shall extend these
coefficients here to include also de Broglie matter waves.

We know that in a real medium (e.g., Bose-Einstein Condensate), it is quite
possible to simulate novelties such as "laboratory optical black holes"
[19-25] or in the acoustic medium "laboratory acoustic black holes" [26-31]
that mimic general relativistic ones. Our methodology here is to follow a
reverse route, viz., start with general relativity, go to effective medium
as an intermediate avatar, therein apply wisdom borrowed from real medium
such as those of Tangherlini, and interpret the effects back in the realm of
general relativity. That is, once we evaluate the non-deterministic
coefficients in the effective medium, we can re-interpret them as chances of
stability of the wormhole that lives in general relativity, and there will
be no more need to talk of the intermediary effective medium.

The purpose of this paper therefore is to revisit the issue of stability
from a physical viewpoint without manifestly needing any mathematical gauge
fixing. We shall explore the extent to which a wormhole throat (Ellis and
phantom) can reflect an ingoing disturbance, caused by the motion of a
photon or a matter particle, observed by asymptotic and near-throat local
observers. Following the argument in [5] quoted above, we shall assume that
the probability of stability is determined by the probability of such
reflection, that is, $R=1$, or equivalently $T=0$, imply certainty of
stability since $R+T=1$. Rephrased in the spirit of [5], intermediate values
of ($R,T$) would mean the degree of uncertainty of reflection, hence
uncertainty of (in)stability. To calculate $R$ and $T$, we shall employ a
combination of two fundamental pre-quantum ideas: Tangherlini's
probabilistic formulation [1] and Hamilton's optical-mechanical analogy
applying it to motion in a gravity field. It turns out that there is a 
\textit{non-zero probability} that Ellis and phantom wormholes could appear
stable depending on the asymptotic speed of the particles as well as on the
location of observers, leading to a new possibility of what we call here
"ghost wormholes" (like ghost stars). We shall propose a thought experiment
for observable deterministic signatures of ghost wormholes.

The paper is organized as follows: We outline in Sec.2, Tangherlini's
formulation dealing with photon motion in real refractive media. Next, in
Sec.3, we implement Hamilton's optical-mechanical analogy in the gravity
field to develop the motion of de Broglie matter waves in the effective
medium. Equations for photon motion follow as a special case. In Sec.4, we
show that the momentum increase of particles entering the real medium has an
exact parallel in the effective medium version of the Pound-Rebka
experiment. We devote Sec.5 to a discussion of Ellis wormhole pointing out
some of its interesting features and provide analytical support to the
possibility of its collapse to Schwarzschild black hole. In Sec.6, we shall
develop Tangherlini's coefficients for the motion of photon and matter de
Broglie waves causing the perturbations. In Sec.7, we apply these
coefficients to Schwarzschild black hole, Ellis wormhole and speculate on
the possibility of ghost wormholes. Sec.8 is devoted to phantom wormholes
and Sec.9 describes a thought experiment. Conclusions are summarized in
Sec.10. We shall take $8\pi G=1$, speed of light in vacuum $c_{0}=1$, unless
specifically restored. Signature convention is ($+---$).

\textbf{2. Tangherlini's formulation}

Tangherlini's considerations tacitly assume scales and clocks unaffected by
the medium and his ingredients are summarized below in (T1)-(T3):

(T1) Light is normally incident on a plane, semi-infinite, homogeneous,
isotropic, non-absorbing real medium. If the photon is transmitted, the
magnitude of its momentum $p^{\prime }$ in the medium is related to the
magnitude of its momentum $p$ in free space by the equation%
\begin{equation}
p^{\prime }=np,
\end{equation}%
where $n$ ($\geq 1$) is the index of refraction. This equation, which holds
independently of the angle of incidence, is an experimental fact in the real
medium. The\ prediction of the increase in momentum in the refractive medium
has been verified to $15\%$ accuracy by early experiments [32]. After a gap
of about half a century, experiments by Jones [33], Jones and Richards [34],
and by Ashkin and Dziedzic [35] confirmed the increase to a remarkable
accuracy.\ Note that the de Broglie relation $p^{\prime }\lambda ^{\prime
}=p\lambda =$ constant (without Planck's constant) together with the
reduction of wavelength in the real optical medium, $\lambda ^{\prime
}=\lambda /n$, which we shall soon verify in the effective medium, lead to
Eq.(1); unprimed quantities\ $p$ and $\lambda $ refer to those in free space
(absence of gravity or effective medium).

From a dynamical standpoint, the treatment of a photon may be based in the
ray approximation on the following Hamiltonian, for negligible dispersion
(see \textit{Eq.(A1)} of Tangherlini [1]):%
\begin{equation}
H^{\prime }=\frac{c_{0}}{n(r)}p^{\prime },
\end{equation}%
where $n=n(r)$ is a slowly varying refractive index for a real medium
oriented so that the boundary is in the transverse plane.

We point out that Eq.(1) with $n=n(r)$ is an experimental fact supported by
the effective medium version of the Pound-Rebka experiment of gravitational
frequency shift (see Sec.4). It thus follows that in the effective medium,
we need not be restricted by real medium constraints such as a constant $n$
or a sharp medium boundary.

(T2) The rate at which energy is transmitted to the real medium is
proportional to the \textit{phase velocity} of classical Hamiltonian
mechanics:%
\begin{equation}
V_{\text{phase}}^{\prime }=\frac{c_{0}}{n(r)}.
\end{equation}%
We shall show that this equation exactly holds in the effective medium as
well and point out two other examples in footnote 4, where central equations
in a real medium resemble those in the effective medium.

(T3) To avoid a deterministic passage of a photon across an impinging
surface, Tangherlini [1] introduced a statistical method, which assumes that
the probability for a photon to be found in a reflected or a transmitted
mode on that surface can be calculated from an \textit{ensemble average}.
The ensemble consists of a large number of identical, widely separated
replica of the media having the same refractive index. There is one ensemble
representative for each incident photon. This condition, which does not
involve Planck' s constant, introduces a kind of pre-quantum indeterminacy
that leaves the motion of individual particles undetermined. The probability
of reflection will be the fraction of the total number of particles that are
observed to have been in the reflected mode, and likewise for the
probability of the transmission.

Let $I$ denote the incident particle flux, let $I_{R}$ be the flux of
reflected particles, and let $I_{T}$ be the flux of transmitted particles
such that 
\begin{equation}
I_{R}+I_{T}=I.
\end{equation}%
In accordance with standard notation, the ratios $R=$ $I_{R}/I$ and $%
T=I_{T}/I$ are taken to define the probabilities of reflection and
transmission respectively. These probabilities satisfy the so-called
conservation of probability condition%
\begin{equation}
R+T=1.
\end{equation}%
Equating the average rate of energy delivered to the ensemble member in
reflected and transmitted modes, Tangherlini [1] derives the coefficients:%
\begin{equation}
R=\frac{(n-1)^{2}}{(n+1)^{2}}\text{, }T=\frac{4n}{(n+1)^{2}},
\end{equation}%
that remain invariant under $n\rightarrow 1/n$. These coefficients exactly
resemble those that are obtained by solving Schr\"{o}dinger equation for a
certain potential [36]. We shall be using Eqs.(6) in the sequel.

\textbf{3. Optical-mechanical analogy: Motion of particles }

We shall implement Hamilton's optical-mechanical analogy in the gravity
field to arrive at the motion of particles and light that exactly reproduces
motions of massive (and massless) particles already otherwise known, for
example, in the Schwarzschild gravity field [10,14]. For our purpose, we
shall consider the generic form of a static, spherically symmetric metric
field in isotropic coordinates 
\begin{equation}
ds^{2}=\Omega ^{2}(\mathbf{r})c_{0}^{2}dt^{2}-\Phi ^{-2}(\mathbf{r}%
)[dr^{2}+r^{2}\left( d\theta ^{2}+\sin ^{2}\theta d\varphi ^{2}\right) ],
\end{equation}%
where $\Omega (\mathbf{r})$ and $\Phi (\mathbf{r})$ are arbitrary metric
functions of spatial coordinates\ $\mathbf{r\equiv (}r,\theta ,\varphi )$ or 
$(x_{1},x_{2},x_{3})$. The coordinate phase speed of light $c^{\prime }(%
\mathbf{r})$, observed by asymptotic observers, is determined by the
condition that the geodesic be null $(ds^{2}=0)$ giving%
\begin{equation}
c^{\prime }(\mathbf{r)=}\left\vert \frac{d\mathbf{r}}{dt}\right\vert
=c_{0}\Phi (\mathbf{r})\Omega (\mathbf{r})=\frac{c_{0}}{n(\mathbf{r})}.
\end{equation}%
Thus the effective index of refraction for light in the gravitational field
is: 
\begin{equation}
n(\mathbf{r})=\Phi ^{-1}\Omega ^{-1}.
\end{equation}%
This index of refraction may be used in any formulation of geometrical
optics using Fermat's principle:

\begin{equation}
\delta \int_{\mathbf{x}_{1}}^{\mathbf{x}_{2}}n\left\vert d\mathbf{r}%
\right\vert =0,
\end{equation}%
where $\delta $ represents a variation of the integral produced by varying
the path between two fixed points $\mathbf{x}_{1}$ and $\mathbf{x}_{2}$ in
the three-dimensional Euclidean space and $\left\vert d\mathbf{r}\right\vert 
$ is the element of the path of integration.

In general relativity, the\ orbits of massive test particles are obtained by
requiring that they be geodesics:

\begin{equation}
\delta \int_{\mathbf{x}_{1},t_{1}}^{\mathbf{x}_{2},t_{2}}m_{0}ds=0,
\end{equation}%
where ($\mathbf{x}_{1},t_{1}$) and ($\mathbf{x}_{2},t_{2}$) are two fixed
points in spacetime, $m_{0}$ is the rest mass of the test particle. Using
Eqs.(7) and (9), the above can be rewritten in the form of Hamilton's
variational principle%
\begin{equation}
\delta \int_{t_{1}}^{t_{2}}L^{\prime }(x_{i},V_{i}^{\prime })dt=0,
\end{equation}%
where the effective Lagrangian is\footnote{%
Here are the two other examples we promised in the preceding section, where
the central equations in a real medium resemble those in the present
effective medium: (i) Starting from the wave equation in a non-uniformly
moving real fluid with refractive index $n$, Leonhardt and Piwnicki [21]
derive the Lagrangian and the Hamiltonian for a light ray as observed by a
lab observer. From the action principle, they arrive at a completely
geometrical picture of ray optics in a moving medium. Light rays are
geodesic lines with respect to the Gordon metric, which, in the comoving
frame of the real fluid element reads: $ds^{2}=\frac{c_{0}^{2}}{n^{2}}%
dt^{2}-d\mathbf{r}^{2}$ and the Lagrangian reads [21]: $L^{\prime
}(x_{i},V_{i}^{\prime })\equiv -m_{0}c_{0}^{2}\times \frac{1}{n}\times \left[
1-\frac{n^{2}V^{\prime 2}}{c_{0}^{2}}\right] ^{1/2}$, where $V^{\prime }$ is
the particle coordinate speed. The Lagrangian for a massive particle in the
comoving frame derived in this paper is [Eq.(13)]: $L^{\prime
}(x_{i},V_{i}^{\prime })\equiv -m_{0}c_{0}^{2}\times \Omega \times \left[ 1-%
\frac{n^{2}V^{\prime 2}}{c_{0}^{2}}\right] ^{1/2}$. Now note that the Gordon
metric with $n$ as the real medium index, can be obtained formally from the
metric (7) above simply by putting $\Phi =1$, $\Omega =\frac{1}{n}$ so that $%
n=\Phi ^{-1}\Omega ^{-1}$. Clearly, the $n$ in Eq.(13) has a different
origin: it derives from the equivalent medium. This notwithstanding, one
finds that the Lagrangians are exactly the same. (ii) The dispersion
relation for light in the comoving frame derived in [21] is $\omega ^{\prime
2}-c_{0}^{2}k^{\prime 2}+(n^{2}-1)\omega ^{\prime 2}=0$. This is precisely
the same as that following from Eq.(69) with $m_{0}=0$ for light. See [16]
for details.} 
\begin{equation}
L^{\prime }(x_{i},V_{i}^{\prime })\equiv -m_{0}c_{0}^{2}\Omega \left[ 1-%
\frac{n^{2}V^{\prime 2}}{c_{0}^{2}}\right] ^{1/2}.
\end{equation}%
In the above expression, the coordinate velocity $V^{\prime }$ is defined by 
$V^{\prime 2}=\sum\limits_{i=1}^{3}V_{i}^{\prime 2}$ and $V_{i}^{\prime
}\equiv \left( \frac{d\mathbf{r}}{dt}\right) $,\textbf{\ }$\mathbf{r}\equiv
(x_{1},x_{2},x_{3})$. Similarly, the canonical coordinate momenta of a
particle in the gravity field are%
\begin{equation}
p_{i}^{\prime }=m_{0}\Omega n^{2}\left[ 1-\frac{n^{2}V^{\prime 2}}{c_{0}^{2}}%
\right] ^{-1/2}V_{i}^{\prime },
\end{equation}%
where the primed notation $p^{\prime }$ is used only to be in conformity
with (T1). The effective Hamiltonian is%
\begin{equation}
H^{\prime }=m_{0}c_{0}^{2}\Omega \left[ 1-\frac{n^{2}V^{\prime 2}}{c_{0}^{2}}%
\right] ^{-1/2},
\end{equation}%
or expressed in terms of momenta%
\begin{equation}
H^{\prime }=m_{0}c_{0}^{2}\left[ \Omega ^{2}+\frac{p^{\prime 2}}{%
n^{2}m_{0}^{2}c_{0}^{2}}\right] ^{1/2},
\end{equation}%
which is a constant of motion. Its value may be calculated from Eq.(15) if
the particle coordinate speed $V^{\prime }$ is known at one point on the
path. Eq.(15) gives the particle coordinate speed in the medium as 
\begin{equation}
V^{\prime }(\mathbf{r})=\frac{c_{0}}{n}\left[ 1-\frac{m_{0}^{2}c_{0}^{4}%
\Omega ^{2}}{H^{\prime 2}}\right] ^{1/2}.
\end{equation}

For the motion of a massive particle inside the medium, we have from
Hamilton's principle, Eq.(12), the Jacobi's form of Maupertuis's principle,
which is, using Eq.(14):%
\begin{equation}
\delta \int_{\mathbf{x}_{1}}^{\mathbf{x}_{2}}p^{\prime }\left\vert d\mathbf{r%
}\right\vert =0,
\end{equation}%
where now the path of integration through three-dimensional space is varied,
subject to conservation of energy, between two fixed points in space, $%
\mathbf{x}_{1}$ and $\mathbf{x}_{2}$, but the times at the end points need
not be held fixed. From Eqs.(14) and (15), we have%
\begin{equation}
p_{i}^{\prime }=H^{\prime }n^{2}V_{i}^{\prime }/c_{0}^{2}.
\end{equation}%
\ \ Specializing to photon motion ($m_{0}=0$), we see that it has the phase
speed [it is also the group speed, see Eq.(72)] in the effective medium 
\begin{equation}
V^{\prime }(\mathbf{r})=\frac{c_{0}}{n},
\end{equation}%
which leads to%
\begin{equation}
H^{\prime }=\frac{c_{0}p^{\prime }}{n},
\end{equation}%
and we have here rederived Tangherlini's Eqs.(2) and (3), as promised. Hence
for a free photon at infinity ($n=1$)%
\begin{equation}
H=c_{0}p.
\end{equation}%
Using the value of $p^{\prime }$ from Eq.(19), Eq.(18) can be written as 
\begin{equation}
\delta \int_{\mathbf{x}_{1}}^{\mathbf{x}_{2}}\frac{H^{\prime }n^{2}V^{\prime
}}{c_{0}^{2}}\left\vert d\mathbf{r}\right\vert =0.
\end{equation}%
From this variational principle follows the exact general-relativistic
equation of motion for a massive particle (see for details, Ref.[14]):%
\begin{equation}
\frac{d^{2}\mathbf{r}}{dA^{2}}=\nabla \left( \frac{1}{2}n^{4}V^{\prime
2}\right) ,
\end{equation}%
which has precisely the form of Newton's second law for the "potential" $-%
\frac{1}{2}n^{4}V^{\prime 2}$. The independent variable $A$ playing the role
of time in Eq.(24) is the optical action, defined by%
\begin{equation}
\left\vert \frac{d\mathbf{r}}{dA}\right\vert =n^{2}V^{\prime },\text{ }dA=%
\frac{dt}{n^{2}}.
\end{equation}%
For photon, using Eq.(20) for $V^{\prime }$, the path Eq.(24) yields%
\begin{equation}
\frac{d^{2}\mathbf{r}}{dA^{2}}=\nabla \left( \frac{1}{2}n^{2}c_{0}^{2}%
\right) .
\end{equation}%
We see that $-\frac{1}{2}n^{2}c_{0}^{2}$ plays the role of "potential" for
the massless particle. Likewise, we can rewrite Eq.(24) as%
\begin{equation}
\frac{d^{2}\mathbf{r}}{dA^{2}}=\nabla \left( \frac{1}{2}N^{2}c_{0}^{2}%
\right) ,
\end{equation}%
such that $N$ ($\equiv n^{2}V^{\prime }/c_{0}$) plays the role of index of
refraction for the massive particle. But since $V^{\prime }=\left\vert d%
\mathbf{r/}dt\right\vert $, it follows that along the trajectory, Eq. (25)
holds. The "potential" on the right hand side and the "acceleration" on the
left in Eqs.(26), (27) are the optical versions describing the mechanical
motion of photon and massive particles respectively.

With the above results at hand, we shall now introduce de Broglie waves of a
massive particle into the medium with index of refraction. In the
geometrical-optics limit, it is possible to speak of a ray, which is the
trajectory of the point-particle, that is, Maupertuis principle for
mechanics (18) is identical with the Fermat's principle for ray optics (10).
This is the essence of Hamilton's \textit{optical-mechanical analogy}
implemented later by de Broglie in his wave-particle duality. The ray will
be the path satisfying Fermat's principle, which we write in the form:%
\begin{equation}
\delta \int_{\mathbf{x}_{1}}^{\mathbf{x}_{2}}k^{\prime }\left\vert d\mathbf{r%
}\right\vert =0,
\end{equation}%
where $k^{\prime }(\mathbf{r)}$\textbf{\ }is the wave number.

Reasoning in exactly the same way as would de Broglie, the paths predicted
by Eq. (23) and Eq. (28) will be the same if the two integrands are the same
functions of the spatial coordinates. However, they may differ by a
multiplicative factor $f$. Thus we require%
\begin{equation}
\frac{H^{\prime }n^{2}V^{\prime }}{c_{0}^{2}}=fk^{\prime }.
\end{equation}%
Going over to the free space limit, the above yields $m_{0}V^{\prime }\gamma
=fk^{\prime }$ and this will be the de Broglie relation for a relativistic
free particle only if $f=\hslash =h/2\pi $, the Planck constant\footnote{%
To be in line with the pre-quantal indeterminacy of Tangherlini, identifying
the constant $f$ with $\hslash $ is not mandatory. However, we have
introduced matter de Broglie waves using $\hslash $ in a semiclassical
manner but there is no role for it as it cancels out in the relevant
expressions that follow [see, e.g., Eq.(37), in which $\hslash $ cancels out
between $p^{\prime }$ and $H^{\prime }$].}. Thus we have the momentum of the
de Broglie waves as 
\begin{equation}
\frac{H^{\prime }n^{2}V^{\prime }}{c_{0}^{2}}=\hslash k^{\prime },
\end{equation}%
which leads to the coordinate wavelength of the de Broglie waves in the
effective medium as%
\begin{equation}
\lambda ^{\prime }=\frac{hc_{0}^{2}}{H^{\prime }n^{2}V^{\prime }}
\end{equation}%
Using Eq.(17) for $V^{\prime }(\mathbf{r)}$, the above can be rewritten as%
\begin{equation}
\lambda ^{\prime }=\frac{hc_{0}}{nH^{\prime }\left[ 1-\frac{%
m_{0}^{2}c_{0}^{4}\Omega ^{2}}{H^{\prime 2}}\right] ^{1/2}},
\end{equation}%
that, in turn, yields%
\begin{equation}
\lambda ^{\prime }n\left[ 1-\frac{m_{0}^{2}c_{0}^{4}\Omega ^{2}}{H^{\prime 2}%
}\right] ^{1/2}=\frac{hc_{0}}{H^{\prime }}=\text{constant.}
\end{equation}

Thus the wave-optics rule for massive particles may be expressed as%
\begin{equation}
\lambda ^{\prime }N=\text{ constant,}
\end{equation}%
where the index of refraction $N$ for the de Broglie waves of massive
particles in the medium is given by%
\begin{equation}
N=n\left[ 1-\frac{m_{0}^{2}c_{0}^{4}\Omega ^{2}}{H^{\prime 2}}\right] ^{1/2}.
\end{equation}%
This can be rewritten as%
\begin{equation}
N=\frac{n^{2}V^{\prime }}{c_{0}}.
\end{equation}%
With the help of Eq.(17), this can be further rewritten as 
\begin{equation}
N=\frac{c_{0}p^{\prime }}{H^{\prime }}.
\end{equation}%
Eqs.(17), (34) and (37) are the generalized equations applicable to matter
de Broglie waves.

Let us specialize to photon motion in the effective medium for which we put $%
m_{0}=0$, $N=n$ in Eqs.(17), (34) and (37) and collect the results in one
place: 
\begin{eqnarray}
V_{\text{phase}}^{\prime } &=&c^{\prime }(r)=\frac{c_{0}}{n(r)}\text{,} \\
\lambda ^{\prime }n(r) &=&\text{constant}=\lambda \text{, } \\
H^{\prime } &=&\frac{c_{0}}{n(r)}p^{\prime }=\text{constant}=c_{0}p.
\end{eqnarray}%
Directly from Eq.(40), we get%
\begin{equation}
p^{\prime }=n(r)p,
\end{equation}%
which is just Tangherlini's Eq.(1) for photon motion. All the above
coordinate relations are the foundation of Tangherlini's formulation [1].
Asymptotic observers measure the index $n$ and photon momentum $p^{\prime }$
connected by Eq.(41) and wavelength $\lambda ^{\prime }$ according to
Eq.(39). The observers \textit{inside} the medium measure proper wave length 
$\widetilde{\lambda }$ defined by $\widetilde{\lambda }=\lambda ^{\prime
}\Phi ^{-1}$, and so Eq.(39) changes to%
\begin{equation}
\widetilde{\lambda }\left( n\Phi \right) =\widetilde{\lambda }\widetilde{n}%
(r)=\text{ constant}=\lambda \text{.}
\end{equation}%
Using de Broglie relations inside the medium, $\widetilde{p}=\hslash /%
\widetilde{\lambda }$ and $p=\hslash /\lambda $ in Eq.(42), when $\hslash $
cancels out, we get%
\begin{equation}
\widetilde{p}=\widetilde{n}(r)p\text{,}
\end{equation}%
where $\widetilde{n}>1$. This is the proper version of Eq.(1) inside the
effective medium\ for ingoing photons that is supported by the Pound-Rebka
experiment as shown below.

\textbf{4. Pound-Rebka experiment}

It is shown above that Tangherlini's Eqs.(1)-(3) are just Eqs.(38), (40),
(41) in the effective medium. The question now is: like the experiments
supporting Eq.(1) in a real optical medium, is there any experiment
confirming the validity of Eq.(43), or the increase of light momentum inside
the effective medium? The answer is yes, and it does not seem widely
recognized that it is just the medium version of the famous Pound-Rebka
frequency shift experiment [37], although this version is purely heuristic
based on the mathematical similitude of the relevant equations.

Let us recall the set-up, where a pulse of light falls to Earth's surface
from the roof of a building of height $h_{0}$. For a Schwarzschild gravity
of mass $M$ describing the Earth's gravity, we have 
\begin{equation}
n_{\text{Sch}}(r)=\frac{\left( 1+\frac{M}{2r}\right) ^{3}}{\left( 1-\frac{M}{%
2r}\right) }\simeq 1+\frac{2M}{r}.
\end{equation}%
Accordingly, using Eq.(41), the coordinate relations (primes replaced by
suffixes for transparency) are%
\begin{equation}
\frac{p_{\text{E}}}{p_{\text{R}}}=\frac{\lambda _{\text{R}}}{\lambda _{\text{%
E}}}=\frac{n_{\text{E}}}{n_{\text{R}}},
\end{equation}%
where the suffix R refers to quantities at the roof, and E refers to those
on the Earth, $r_{\text{E}}$ is the radius of the Earth. (The isotropic
radius $r_{\text{E}}$ is far greater than Earth's black hole radius $\frac{GM%
}{2c_{0}^{2}}$ of a few centimeters, so Earth's radius in "standard"
coordinates would be very nearly the same as $r_{\text{E}}$, differing by
those few centimeters). Note that both the events of emission and reception
are taking place inside the effective medium so that, using Eqs.(42) and
(43), we get 
\begin{equation}
\frac{\widetilde{p}_{\text{E}}}{\widetilde{p}_{\text{R}}}=\frac{\widetilde{%
\lambda }_{\text{R}}}{\widetilde{\lambda }_{\text{E}}}=\frac{n_{\text{E}%
}\Phi _{\text{E}}}{n_{\text{R}}\Phi _{\text{R}}},
\end{equation}%
where $n\Phi >1$. Neglecting $\left( h_{0}/r_{\text{E}}\right) ^{2}$ and
higher orders, we get%
\begin{eqnarray}
\frac{n_{\text{E}}}{n_{\text{R}}} &\simeq &\left( 1+\frac{2M}{r_{\text{E}}}%
\right) /\left( 1+\frac{2M}{r_{\text{E}}+h_{0}}\right) \simeq 1+\frac{2M}{r_{%
\text{E}}^{2}}h_{0}, \\
\frac{\Phi _{\text{E}}}{\Phi _{\text{R}}} &\simeq &1-\frac{M}{r_{\text{E}%
}^{2}}h_{0}\text{.}
\end{eqnarray}%
Hence, to first order in $M$,%
\begin{equation}
\frac{\widetilde{p}_{\text{E}}}{\widetilde{p}_{\text{R}}}=\frac{\widetilde{%
\lambda }_{\text{R}}}{\widetilde{\lambda }_{\text{E}}}=1+\frac{M}{r_{\text{E}%
}^{2}}h_{0}.
\end{equation}%
The second equality has been very accurately confirmed by the Pound-Rebka
experiment in gravity, which now implies that $\widetilde{p}_{\text{E}}>$ $%
\widetilde{p}_{\text{R}}$. A similar experiment with matter de Broglie waves
testing the equation%
\begin{equation}
\frac{\widetilde{p}_{\text{E}}}{\widetilde{p}_{\text{R}}}=\frac{\widetilde{%
\lambda }_{\text{R}}}{\widetilde{\lambda }_{\text{E}}}=\frac{N_{\text{E}%
}\Phi _{\text{E}}}{N_{\text{R}}\Phi _{\text{R}}},
\end{equation}%
where $N\Phi >1$, has also been suggested in the literature [14].

Therefore, what is known as the gravitational frequency shift could be
interpreted as the momentum increase of the ingoing pulse in the effective
medium. One sees that the effective medium behaves exactly like a real
medium in this regard. The premise for the application of Tangherlini's
approach to wormholes is now ready at hand. Accordingly, in the next section
we shall briefly discuss Ellis wormhole and thereafter develop the
reflection and transmission coefficients for photon and de Broglie waves.
Later we shall apply them also to phantom wormhole.

\textbf{5. Ellis wormhole and Schwarzschild black hole}

This wormhole is a solution of Einstein's field equations%
\begin{equation}
R_{\mu \nu }=\varepsilon \phi _{,\mu }\phi _{,\nu }
\end{equation}%
where $\phi _{,\mu }\equiv \frac{\partial \phi }{\partial x^{\mu }}$, $\phi $
being a massless scalar field. Choosing $\varepsilon =-1$, which represents
the ghost regime, the static spherically symmetric massive Ellis wormhole in
isotropic coordinates can be written as%
\begin{equation}
d\tau ^{2}=\Omega ^{2}(r)dt^{2}-\Phi ^{-2}(r)[dr^{2}+r^{2}\left( d\theta
^{2}+\sin ^{2}\theta d\varphi ^{2}\right) ],
\end{equation}%
where

\begin{align}
\Omega ^{2}(r)& =\exp \left[ 2\epsilon +4\gamma \tan ^{-1}(r/B)\right] ,%
\text{ } \\
\Phi ^{-2}(r)& =\left( 1+\frac{B^{2}}{r^{2}}\right) ^{2}\exp \left[ 2\zeta
-4\gamma \tan ^{-1}(r/B)\right] , \\
\phi (r)& =4\lambda \tan ^{-1}(r/B)\text{, }
\end{align}%
with the constraint $2\lambda ^{2}=1+\gamma ^{2}$ following from the field
Eq.(51), $B$ and $\gamma $ being arbitrary constants. Asymptotic flatness
requires that $\epsilon =-\pi \gamma $ and $\zeta =\pi \gamma $. The
spacetime has no horizon, is twice asymptotically flat and is a traversable,
everywhere regular natural wormhole (as opposed to artificially assembled
ones) with the throat appearing at the isotropic coordinate radius $r_{\text{%
th}}=\frac{M}{2\gamma }\left[ \gamma +\sqrt{1+\gamma ^{2}}\right] $.

To put it into a familiar form, transform the radial coordinate $%
r\rightarrow \ell $ by $\ell =r-\frac{B^{2}}{r}$, where $\ell \in (-\infty
,\infty )$ corresponding to $r\in (0,\infty )$. Redefining $B=m/2$, and
using the identity 
\begin{equation}
2\tan ^{-1}\left( \frac{\ell +\sqrt{\ell ^{2}+m^{2}}}{m}\right) \equiv \frac{%
\pi }{2}+\tan ^{-1}\left( \frac{\ell }{m}\right) ,
\end{equation}%
we see that the solution set (53)-(55) transforms exactly to the form of
Ellis class III wormhole solution (\textit{the} Ellis wormhole) [2]:

\begin{align}
d\tau _{\text{Ellis}}^{2}& =Fdt^{2}-F^{-1}[d\ell ^{2}+(\ell
^{2}+m^{2})(d\theta ^{2}+\sin ^{2}\theta d\varphi ^{2})] \\
F& =\exp \left[ -\pi \gamma +2\gamma \tan ^{-1}\left( \frac{\ell }{m}\right) %
\right] \\
\phi & =\lambda \left[ \pi +2\tan ^{-1}\left( \frac{\ell }{m}\right) \right]
,\text{ \ }
\end{align}%
with the constraint $2\lambda ^{2}=1+\gamma ^{2}$ remaining the same. Note
that $\ell $ is not the proper radial distance, the latter being given by $%
L=\int F^{-1/2}d\ell $. The wormhole has manifestly two asymptotically flat
regions, one with positive mass $M$ ($=m\gamma )$ and the other with
negative mass $-Me^{\pi \gamma }$, on either side of a regular throat
(minimal area radius) at $\ell _{\text{th}}=M$. A single extended coordinate
chart $(t,\ell ,\theta ,\varphi )$ across the regular throat now covers both
the sides. The same solution is used in [5] that can be seen by renaming $%
\ell \equiv x$, $m\equiv b$ and their constants as $\gamma _{1}=\gamma $, $%
\gamma _{0}=-\frac{\pi }{2}\gamma $, $\Phi _{0}=\lambda \pi $, $\Phi
_{1}=2\lambda $. The constraint $2\lambda ^{2}=1+\gamma ^{2}$ is just their
equation: $\Phi _{1}^{2}=2(1+\gamma _{1}^{2}).$

Using the Ellis wormhole metric functions $\Omega $ and $\Phi $ from
Eqs.(53),(54) and identifying the constant as $B=\frac{M}{2\gamma }$, we see
that the equivalent refractive index for photon motion perceived by
asymptotic observers is%
\begin{eqnarray}
n(r) &=&\Phi ^{-1}\Omega ^{-1}=\left( 1+\frac{M^{2}}{4r^{2}\gamma ^{2}}%
\right) \exp \left[ 2\pi \gamma -4\gamma \tan ^{-1}\left( \frac{2r\gamma }{M}%
\right) \right] \\
&\simeq &1+\frac{2M}{r}+\left( 2+\frac{1}{4\gamma ^{2}}\right) \left( \frac{M%
}{r}\right) ^{2}+\left( \frac{1+4\gamma ^{2}}{3\gamma ^{2}}\right) \left( 
\frac{M}{r}\right) ^{3}+...
\end{eqnarray}%
where $M=m\gamma $ is the Keplerian mass observed by the asymptotic
observers. This expansion does not coincide with the expansion of refractive
index for the Schwarzschild gravity for the same mass $M$, viz., 
\begin{equation}
n_{\text{Sch}}(r)\simeq 1+\frac{2M}{r}+\left( \frac{7}{4}\right) \left( 
\frac{M}{r}\right) ^{2}+\left( \frac{M}{r}\right) ^{3}+...
\end{equation}%
for any \textit{real} $\gamma $. Here we recall that a collapse to a
Schwarzschild black hole is possible from the Ellis class I solution [2]
(naked singularity, not a wormhole) because there is a smooth passage from
it to the black hole under the trivial choice $\gamma =1$ (the so called
no-hair theorem). In its vacuum Brans-Dicke representation of the Ellis
class I solution, the scalar field is radiated away in the
Oppenheimer-Snyder collapse resulting into a black hole identical to that in
general relativity, as has been shown by Scheel, Shapiro and Teukolsky [38].

The passage from the Ellis wormhole to Schwarzschild black hole happens 
\textit{only} if one puts $\gamma =-i$ in Eq.(61). Therefore, the collapse
to the latter might at best seem fortuitous. That however need not be the
case although the route is by no means trivial. Consider the metric (53-55)
under inversion and Wick-like rotation: $r\rightarrow \frac{1}{r}$, $%
B\rightarrow \frac{i}{B}$, $\gamma =-i.$ With $B=\frac{M}{2\gamma }$, we
obtain $\Omega ^{2}(r)=e^{-4\tanh ^{-1}\left( \frac{M}{2r}\right) }$. Using
further the identity $e^{\tanh ^{-1}(x)}\equiv \frac{1}{2}\ln \left( \frac{%
1+x}{1-x}\right) $, we get $\Omega ^{2}(r)=\left( \frac{1-\frac{M}{2r}}{1+%
\frac{M}{2r}}\right) ^{2}$ and $\Phi ^{-2}(r)=\left( 1+\frac{M}{2r}\right)
^{4}$, which is just the Schwarzschild metric in the isotropic form. And of
course the scalar field $\phi =0$, as it should be, and the isotropic
coordinate radius of the throat $r_{\text{th}}=\frac{M}{2\gamma }\left[
\gamma +\sqrt{1+\gamma ^{2}}\right] $ converts to coordinate horizon: $r_{%
\text{hor}}=\frac{M}{2}$. This derivation provides a completely analytic
reason as to why Gonz\'{a}lez, Guzm\'{a}n and Sarbach [6] get a "strong
indication" in their numerical simulation of the Ellis wormhole collapsing
to the Schwarzschild black hole.

\textbf{6. Reflection and Transmission coefficients}

\textit{(a) Light pulse:}

The throat of the Ellis wormhole occurs at the minimum of the areal radius $%
R(r)=r\Phi ^{-1}(r)$, which gives 
\begin{equation}
r_{\text{th}}=\frac{M}{2\gamma }\left[ \gamma +\sqrt{1+\gamma ^{2}}\right] .
\end{equation}%
The coefficients for the ingoing photon pulse at the throat as observed by 
\textit{asymptotic observers (a.o.) }are given, using Eqs.(6) and (60) with $%
n$, by: 
\begin{eqnarray}
R_{\text{photon}}^{\text{a.o.}} &=&\frac{\left[ n(r_{\text{th}})-1\right]
^{2}}{\left[ n(r_{\text{th}})+1\right] ^{2}}\text{, } \\
T_{\text{photon}}^{\text{a.o.}} &=&\frac{4n(r_{\text{th}})}{\left[ n(r_{%
\text{th}})+1\right] ^{2}}.
\end{eqnarray}%
The coefficients for the ingoing photon pulse at the throat as observed by%
\textit{\ near-throat local observers (l.o.)} are given, using Eqs.(6) and
(60) with $\widetilde{n}$ ($=n\Phi $) by: 
\begin{eqnarray}
\widetilde{R}_{\text{photon}}^{\text{l.o.}} &=&\frac{\left[ \widetilde{n}(r_{%
\text{th}})-1\right] ^{2}}{\left[ \widetilde{n}(r_{\text{th}})+1\right] ^{2}}%
, \\
\widetilde{T}_{\text{photon}}^{\text{l.o.}} &=&\frac{4\widetilde{n}(r_{\text{%
th}})}{\left[ \widetilde{n}(r_{\text{th}})+1\right] ^{2}}.
\end{eqnarray}

\textit{(b) de Broglie matter waves:}

The phase velocity of de Broglie waves is $v_{\text{phase}}^{\prime }=\frac{%
\omega ^{\prime }}{k^{\prime }}=\frac{H^{\prime }}{p^{\prime }}$, or, with
use of Eq.(37),%
\begin{equation}
v_{\text{phase}}^{\prime }=\frac{c_{0}}{N},
\end{equation}%
a relation which demonstrates once again that $N$ is playing its desired
role as an index of refraction for de Broglie matter waves. Inserting $%
p^{\prime }=\hslash k^{\prime }$ and $H^{\prime }=\hslash \omega ^{\prime }$
into the square of Eq.(16), we obtain%
\begin{equation}
\hslash ^{2}\omega ^{\prime 2}=m_{0}^{2}c_{0}^{4}\Omega ^{2}+\frac{%
c_{0}^{2}\hslash ^{2}k^{\prime 2}}{n^{2}}.
\end{equation}%
Differentiating with respect to $k^{\prime }$, we obtain%
\begin{equation}
\frac{\omega ^{\prime }}{k^{\prime }}\frac{d\omega ^{\prime }}{dk^{\prime }}=%
\frac{c_{0}^{2}}{n^{2}}\text{, or, }v_{\text{phase}}^{\prime }v_{\text{group}%
}^{\prime }=\frac{c_{0}^{2}}{n^{2}},
\end{equation}%
where $v_{\text{group}}^{\prime }=\frac{d\omega ^{\prime }}{dk^{\prime }}$
is the group velocity of the de Broglie waves\footnote{%
Observers situated at infinity see that near the horizon, $n\rightarrow
\infty $ so that $v_{\text{phase}}^{\prime }$, $v_{\text{group}}^{\prime
}\rightarrow 0$, both for light and matter de Broglie waves, in perfect
accordance with general relativistic scenario. It is exactly here that the
conditions for laboratory "optical black holes" required by Leonhardt and
Piwnicki [21,22] and Hau \textit{et al.} [20] are met most naturally, that
is, extremely low group velocity or high refractive index. Group velocity of
light as low as $17$ meters/sec in Bose-Einstein condensate has been
attained in the laboratory [20]. See also [23]. In this respect, optical and
gravitational black holes look much similar.
\par
{}}. Using Eq.(70), we get%
\begin{equation}
v_{\text{group}}^{\prime }=\frac{c_{0}N}{n^{2}}.
\end{equation}%
For photons, $N=n$ and so 
\begin{equation}
V_{\text{phase}}^{\prime }=V_{\text{group}}^{\prime }=\frac{c_{0}}{n}.
\end{equation}%
This is the same as Eq.(38). Substituting the expression for $N$ from Eq.
(36) in Eq.(71), we see that%
\begin{equation}
v_{\text{group}}^{\prime }=V^{\prime }.
\end{equation}%
Thus, $V^{\prime }$ defined by Eq.(17) is the group velocity of de Broglie
matter waves. In the asymptotic limit $r\rightarrow \infty $, we have $n=1$,
and the free particle has $V^{\prime }=$ $V=$ constant so that from Eq.(36),
we find the generic result%
\begin{equation}
N_{\infty }=\frac{V}{c_{0}}\equiv \beta .
\end{equation}%
Using $H^{\prime }=h\nu ^{\prime }$, $\lambda ^{\prime }\nu ^{\prime
}=c^{\prime }=\frac{c_{0}}{n}$, and introducing the Compton wavelength of
the particle by $\lambda _{\text{C}}=\frac{h}{m_{0}c_{0}}$, we can rewrite
Eq.(35) as 
\begin{equation}
N(r)=n(r)\left[ 1-\left( \frac{\lambda ^{\prime }}{\lambda _{\text{C}}}%
\right) ^{2}\Phi ^{-2}(r)\right] ^{1/2}\text{.}
\end{equation}%
From Eq.(34), we get%
\begin{equation}
\lambda ^{\prime }N=\lambda N_{\infty }\text{.}
\end{equation}%
Using the de Broglie relations $p^{\prime }=\hslash /\lambda ^{\prime }$ and 
$p=\hslash /\lambda $ in Eq.(76), where $\hslash $ cancels out, we get%
\begin{equation}
p^{\prime }=\frac{N(r)}{\beta }p=\overline{N}(r)p,\text{ }\overline{N}(r)=%
\frac{N(r)}{\beta }>1,
\end{equation}%
at $r\geq r_{\text{th}}$ for wormholes. This $\overline{N}$ generalizes
Eq.(41) to massive particles, where $c^{\prime }=c_{0}/\overline{N}(r)$. The
expressions of the coefficients for the ingoing de Broglie waves as observed
by \textit{asymptotic observers (a.o.) }now become

\begin{eqnarray}
\overline{R}_{\text{dB}}^{\text{a.o.}} &=&\frac{\left[ \overline{N}(r_{\text{%
th}})-1\right] ^{2}}{\left[ \overline{N}(r_{\text{th}})+1\right] ^{2}}\text{%
, } \\
\overline{T}_{\text{dB}}^{\text{a.o.}} &=&\frac{4\overline{N}(r_{\text{th}})%
}{\left[ \overline{N}(r_{\text{th}})+1\right] ^{2}}.
\end{eqnarray}

The observers inside the medium measure proper wave length $\widetilde{%
\lambda }$ defined by $\widetilde{\lambda }=\lambda ^{\prime }\Phi ^{-1}$,
and so Eq.(76) changes to%
\begin{equation}
\widetilde{\lambda }\left( \Phi N\right) =\widetilde{\lambda }\widetilde{N}%
(r)=\text{ constant}=\lambda N_{\infty }\text{,}
\end{equation}%
where%
\begin{equation}
\widetilde{N}(r)=\Phi (r)N(r)=\Phi (r)n(r)\left[ 1-\left( \frac{\widetilde{%
\lambda }}{\lambda _{\text{C}}}\right) ^{2}\right] ^{1/2}.
\end{equation}%
Using de Broglie relations inside the medium, $\widetilde{p}=\hslash /%
\widetilde{\lambda }$ and $p=\hslash /\lambda $ in Eq.(80), where again $%
\hslash $ cancels out, we get%
\begin{equation}
\widetilde{p}=\frac{\widetilde{N}(r)}{\beta }p=\widehat{N}(r)p\text{, }%
\widehat{N}(r)=\frac{\widetilde{N}(r)}{\beta }=\frac{\Phi (r)N(r)}{\beta }>1,
\end{equation}%
at $r\geq r_{\text{th}}$ for wormholes. This is the proper language version
of Eq.(1) in the effective medium\ for ingoing massive particles.
Accordingly, the expressions of the coefficients for the ingoing de Broglie
waves as observed by \textit{near-throat local observers (l.o.)} now become

\begin{eqnarray}
\widehat{R}_{\text{dB}}^{\text{l.o}} &=&\frac{\left[ \widehat{N}(r_{\text{th}%
})-1\right] ^{2}}{\left[ \widehat{N}(r_{\text{th}})+1\right] ^{2}}\text{, }
\\
\widehat{T}_{\text{dB}}^{\text{l.o.}} &=&\frac{4\widehat{N}(r_{\text{th}})}{%
\left[ \widehat{N}(r_{\text{th}})+1\right] ^{2}}.
\end{eqnarray}

\textit{Four pairs of equations (64,65), (66,67) and (78,79), (83,84) are
the coefficients that we have been looking for.} Looking at them, we see
that except for the first pair, the formulation of other three pairs is of
hybrid nature as they have one leg in the effective medium defined by the
index $n$ and the other in the metric function $\Phi $. For our purposes, it
is enough to treat all of the equation pairs as merely some functions of $r$
and evaluate the coefficients for different observers~in different metrics.

For applying the last two pairs of formulas, it is necessary to compare the
two wavelengths $\lambda ^{\prime }$ and $\lambda _{\text{C}}$ appearing in
them. The coordinate de Broglie wavelength $\lambda ^{\prime }$ is related
to a particle's momentum $p^{\prime }$, while the Compton wavelength $%
\lambda _{\text{C}}$ is based on its rest mass $m_{0}$ $-$ it is the
wavelength of a photon whose energy is the same as the rest-mass energy of
the particle. Thus, the de Broglie wavelength of a particle at rest is
infinite, not the Compton wavelength. Highly non-relativistic particles can
show wave properties over scales much larger than their Compton wavelength.
A better way to think of the comparison is that when the coordinate de
Broglie wavelength $\lambda ^{\prime }$ or proper wavelength $\widetilde{%
\lambda }$ becomes shorter than the Compton wavelength $\lambda _{\text{C}}$%
, relativistic effects become very important.

Defining $\lambda ^{\prime }/\lambda _{\text{C}}=\alpha $, we can rewrite 
\begin{equation}
N(r)=n(r)\left[ 1-\left( \frac{\alpha }{\Phi }\right) ^{2}\right] ^{1/2}.
\end{equation}%
To preserve a real $N(r)$, we should maintain the condition that%
\begin{equation}
\alpha \leq \Phi (r)\text{ }\forall r.
\end{equation}%
Equality implies that $N(r)=0$, which in turn suggests that $v_{\text{phase}%
}^{\prime }=v_{\text{group}}^{\prime }=\infty $, an unphysical result
violating special relativity. Therefore, $\alpha <\Phi (r)$. Likewise, for $%
\widetilde{N}(r)$ in Eq.(81), we must maintain $\widetilde{\lambda }<\lambda
_{\text{C}}$.

\textbf{7. Applications}

\textit{(i) Schwarzschild black hole}

In this case, $r_{\text{th}}$ should be replaced by $r_{\text{hor}}=M/2$.
For ingoing photons, the indices are $n_{\text{Sch}}(r)=$ $\frac{\left( 1+%
\frac{M}{2r}\right) ^{3}}{\left( 1-\frac{M}{2r}\right) }$ and $\widetilde{n}%
_{\text{Sch}}(r)=\frac{1+\frac{M}{2r}}{1-\frac{M}{2r}}$ such that $n_{\text{%
Sch}}(r)$, $\widetilde{n}_{\text{Sch}}(r)\rightarrow \infty $ at the
horizon. Hence, the coefficients at the horizon as would appear to different
observers are: $R_{\text{photon}}^{\text{a.o.}}=\widetilde{R}_{\text{photon}%
}^{\text{l.o.}}=1,T_{\text{photon}}^{\text{a.o.}}=\widetilde{T}_{\text{photon%
}}^{\text{l.o.}}=0$. The reflexivity is unity, hence the horizon is a stable
surface, which is also independently known to be true. But if such
reflection back into space really takes place, a black hole will not be a
black hole any more!

This however would not happen because the time $t$ needed for the pulse to
emerge from $r_{\text{hor}}$ and reach the asymptotic observer is infinite
as the integrations $t=\int_{r_{\text{hor}}}^{\infty }\frac{n_{\text{Sch}}(r)%
}{c_{0}}dr$ and $\widetilde{t}=\int_{r_{\text{hor}}}^{\infty }\frac{%
\widetilde{n}_{\text{Sch}}(r)}{c_{0}}dr$ show. Light from the horizon will
never reach the asymptotic observer justifying the name black hole.
Therefore, there will be no true reflection in the ordinary sense $-$ it has
to be understood only by \textit{complementarity} in the sense that the
transmittivity coefficient is zero making up $R+T=1$, that is, light cannot
transmit across the horizon either, so that the surface $r_{\text{hor}}=M/2$
remains stable\footnote{%
No reflection, no transmission, where does all the light then go? For the
Schwarzschild black hole of mass $M$, $n_{\text{Sch}}\rightarrow \infty $
and $c^{\prime }=0$, as the horizon $r_{\text{hor}}=M/2$ is approached by
incoming light from infinity. Asymptotic observers thus see light standing
"still" at $r_{\text{hor}}$, which is consistent with $T_{\text{photon}}^{%
\text{a.o.}}=0$ occurring exactly at $r_{\text{hor}}$. Similarly, local
observers situated very near $r_{\text{hor}}$, on the other hand, measure
proper quantities, $\widetilde{n}_{\text{Sch}}\rightarrow \infty $, again
yielding $\widetilde{T}_{\text{photon}}^{\text{l.o.}}=0$. These observers
would see light or particles never transmitting across the horizon but
moving (as opposed to standing still) at the speed of light, $\widetilde{c}(%
\mathbf{r})=\left\vert \frac{dL}{d\tau }\right\vert =c_{0}$, right on the
horizon making it a "light sphere" (see, e.g., Ref.[39]) in conformity with
general relativity. To either type of observers, asymptotic or local,
complete lack of transmission across horizon is the reason for stability. We
thus have an alternative description of the stable horizon, usually
discussed within the perturbation scheme to the full spacetime fabric. The
effective medium approach has no conflict with the usual geometric approach
of general relativity, as also exemplified in Sec.4.
\par
{}
\par
{}}. In the free space, $r\rightarrow \infty $, we have $n_{\text{Sch}}(r)=1$
and so $R_{\text{photon}}^{\text{a.o.}}=0$ and $T_{\text{photon}}^{\text{a.o.%
}}=1$, that is, photons are fully transmitted, as expected.

For ingoing de Broglie waves with asymptotic speed $V$, the indices, using
Eqs.(75) and (77), are 
\begin{equation}
\overline{N}_{\text{Sch}}(r)=\frac{n_{\text{Sch}}(r)}{\beta }\left[ 1-\alpha
^{2}\left( 1+\frac{M}{2r}\right) ^{4}\right] ^{1/2}.
\end{equation}%
Here again, using Eqs.(75),(77) and (81), we get $\overline{N}_{\text{Sch}%
}(r),\widehat{N}_{\text{Sch}}(r)\rightarrow \infty $ at the horizon, since $%
n_{\text{Sch}}(r)$, $\widetilde{n}_{\text{Sch}}(r)\rightarrow \infty $
there. Hence, it follows that $\overline{R}_{\text{dB}}^{\text{a.o.}}=%
\widehat{R}_{\text{dB}}^{\text{l.o.}}=1$, $\overline{T}_{\text{dB}}^{\text{%
a.o.}}=\widehat{T}_{\text{dB}}^{\text{l.o.}}=0$, independently of the value
of $V$ or the location of observers. However, even though the reflexivity is
unity, time for the reflected de Broglie waves to reach asymptotic observer
is infinite as $\overline{N}_{\text{Sch}}(r)\rightarrow \infty $. Hence,
like in the case of light motion discussed above, since $\overline{t}%
=\int_{r_{\text{hor}}}^{\infty }\frac{\overline{N}_{\text{Sch}}(r)}{c_{0}}dr$
and $\widehat{t}=\int_{r_{\text{hor}}}^{\infty }\frac{,\widehat{N}_{\text{Sch%
}}(r)}{c_{0}}dr$ is infinite, there will be no true reflection of matter
particles and stability of the horizon is to be understood again only by the
total lack of transmittivity across it.

In free space, at $r\rightarrow \infty $, using Eqs.(74) and (77), we get $%
\overline{N}_{\text{Sch}}(\infty )=\frac{N_{\infty }}{\beta }=1$. Using
Eqs.(78) and (79), we then find $\overline{R}_{\text{dB}}^{\text{a.o.}}=0$
and $\overline{T}_{\text{dB}}^{\text{a.o.}}=1$, implying complete
transmission of de Broglie waves into the medium, as expected. Hence, the
Schwarzschild horizon is always a stable surface, no matter whether the
perturbing pulse is a photon or a de Broglie wave.

\textit{(ii) Massive Ellis wormhole}

The effective indices $n(r)$ and $N(r)$ are given in Eqs.(60) and (75)
respectively and the isotropic throat radius is given by $r_{\text{th}}=%
\frac{M}{2\gamma }\left[ \gamma +\sqrt{1+\gamma ^{2}}\right] $ with $\alpha
<\Phi (r_{\text{th}})$ as required for the reality of $N(r)$. It is seen
that, as $\gamma \rightarrow \infty $, $r_{\text{th}}\rightarrow M$. This is
the minimum of $r_{\text{th}}$, while in general, $r_{\text{th}}>M$ for
finite $\gamma >0$ . Also, Fig.1 shows $\Phi (r_{\text{th}})$ vs $r_{\text{th%
}}$ for different $\gamma $ so that the required constraint $\alpha <\Phi
(r_{\text{th}})$ can be used, while choosing the value of $\alpha $. With
these information at hand, choosing units such that $M=1$, and for a range
of real non-zero values of $\gamma $, we graphically study the four pairs of
coefficients.

For ingoing photons, Fig.2 displays the values of coefficients: The first
thing to note is that the values of ($R,T$) become nearly independent of the
value of $\gamma $ for large $\gamma $ or at the minimum of throat $r_{\text{%
th}}$. The plot shows the perception of asymptotic observers,\textit{\ }to
whom the values appear to be $R_{\text{photon}}^{\text{a.o.}}\simeq $ $0.60$
and $T_{\text{photon}}^{\text{a.o.}}\simeq 0.40$ [Eqs.(64,65)] independently
of the size of the throat. This means\ that for those observers\textit{\ }%
there is a good probability that the ingoing photon will be reflected back
from the throat providing chances of stability of the Ellis wormhole%
\footnote{%
Ordinary reflection is possible here because a wormhole throat is not a
black hole horizon.}. On the other hand, near-throat\textit{\ local observers%
} observe the index as $\widetilde{n}$, which yields the coefficients $%
\widetilde{R}_{\text{photon}}^{\text{l.o.}}\simeq $ $0.20$ and $\widetilde{T}%
_{\text{photon}}^{\text{l.o.}}\simeq 0.80$ [Eqs.(66,67)] independently of
the size of the throat. The probability of reflection is evidently very
small meaning that the local observers are most likely to observe
instability of the wormhole. The situation here is quite unlike the black
hole horizon since at the throat, effective refractive indices $n(r)$ and $%
N(r)$ don't diverge.

The intriguing thing is that the probabilities and times of observations
could conspire in such a way that, while the near-throat local observer
finds low probability of reflection ($\widetilde{R}_{\text{photon}}^{\text{%
l.o.}}\simeq $ $0.20$), meaning collapse (instability) of the wormhole, the
asymptotic observer finds higher probability of reflection ($R_{\text{photon}%
}^{\text{a.o.}}\simeq $ $0.60$), meaning that he is more likely to find the
wormhole living (stability). This gives rise to a situation very similar to,
but not exactly the same as, that of a ghost star. Since the Ellis wormhole
spacetime including the throat surface is regular, the information of life
prior to collapse would travel to asymptotic observers, who would then see
it as a living wormhole consistent with higher probabilities for reflection
observed by them. One could say in this case that the asymptotic observer is
seeing a \textit{ghost wormhole} (nothing to do with the ghost scalar field, 
$\varepsilon =-1$).

For ingoing de Broglie waves as perceived by asymptotic observers, we choose
as an example, $\alpha =0.01<\Phi (r_{\text{th}})$ and $\beta =0.2$. Fig.3.
shows that\ $\overline{R}_{\text{dB}}^{\text{a.o.}}$reaches as high a value
as $0.76$, while $\overline{T}_{\text{dB}}^{\text{a.o.}}$\ reaches a value $%
0.24$\ independently of the value of $\gamma $\ or the size of the throat.
In fact, as $\beta \rightarrow 0$, it follows that $\overline{N}\rightarrow
\infty $, so that $\overline{R}_{\text{dB}}^{\text{a.o.}}\rightarrow 1$, $%
\overline{T}_{\text{dB}}^{\text{a.o.}}\rightarrow 0$, leading to total
reflection, hence stability, in this limit. The same figure also shows that
the higher the value of $\beta $, say $0.5$, the lower is the probability of
reflection $\overline{R}_{\text{dB}}^{\text{a.o.}}$, that is, the matter
particles are more likely to penetrate the throat causing instability. This
is a physically acceptable result.

Fig.4. shows the coefficients for ingoing de Broglie waves in the Ellis
wormhole as perceived by the near-throat local observers. Plot shows de
Broglie waves of high and low velocities $\beta =0.9$ and $0.01$. The lesser
the velocity, the higher is the $\widehat{R}_{\text{dB}}^{\text{l.o.}}$.
When $\beta \rightarrow 0$ (low velocity perturbation at asymptotics) ,
stability is achieved: $\widehat{N}\ \rightarrow \infty ,\widehat{R}_{\text{%
dB}}^{\text{l.o.}}\rightarrow 1$, $\widehat{T}_{\text{dB}}^{\text{l.o.}%
}\rightarrow 0$. However, stability of the wormhole is still not a certainty
except in the limit $\beta \rightarrow 0$, because\ $R_{\text{photon}}^{%
\text{a.o.}}$, $\overline{R}_{\text{dB}}^{\text{a.o.}}$ or $\widehat{R}_{%
\text{dB}}^{\text{l.o.}}$ etc are not exactly unity, unlike in the case of
Schwarzschild horizon. By the same token, instability is not a certainty
either, when statistical arguments are used.

\textit{(iii) Massless Ellis wormhole}

This is the most discussed wormhole, which is often used also as a good
exemplar in the literature [40]. This is obtained from Eq.(47) by setting
the Keplerian mass $M=m\gamma =0$. This value can be attained by two
variants $-$ either by setting $m=0,\gamma \neq 0$, which leads to a trivial
flat spacetime, or by setting $m\neq 0$, $\gamma =0$, which leads to what
one might call a zero (Keplerian) mass Ellis wormhole. The two mouths
nonetheless have non-zero ADM masses $+m/2$ and $-m/2$ that add to zero.
These masses are formed entirely by the scalar field $\phi (r)=2\sqrt{2}\tan
^{-1}\left( \frac{2r}{m}\right) $, as the integration of $\phi $ on two
sides will show, and are responsible for gravitational microlensing, as
worked out in detail by Abe [41].

For ingoing photon, we first note that, using Eq.(60) for $n(r)$, putting in
it $M=m\gamma $, and choosing $\gamma =0$ in the resulting expression, we
find 
\begin{equation}
n(r)=1+\frac{m^{2}}{4r^{2}},
\end{equation}%
and the throat, under similar choice, occur at a radius 
\begin{equation}
r_{\text{th}}=\frac{m}{2}\left[ \gamma +\sqrt{1+\gamma ^{2}}\right]
\rightarrow \frac{m}{2}.
\end{equation}%
This gives, at the throat, $n(r_{\text{th}})=2$ and for asymptotic
observers, from Eqs.(64,65), the exact values of the coefiicients 
\begin{equation}
R_{\text{photon}}^{\text{a.o.}}=\frac{1}{9}\text{, }T_{\text{photon}}^{\text{%
a.o.}}=\frac{8}{9}.
\end{equation}%
An interesting result is that the near-throat local observers will notice $%
\widetilde{n}(r)=n\Phi =1$ (since $n=\Phi ^{-1}$), and consequently, from
Eqs.(66,67), we get the exact values%
\begin{equation}
\widetilde{R}_{\text{photon}}^{\text{l.o.}}=0\text{, }\widetilde{T}_{\text{%
photon}}^{\text{l.o.}}=1,
\end{equation}%
which means that the photon will mimic its motion like in free space and
completely transmit across the throat. The wormhole will appear definitively
unstable to the local observers. Eqs.(90) and (91) together lead to a
genuine possibility of zero mass ghost wormhole. This makes sense as Abe
[41] has worked out lensing signatures of the living wormhole. Since the
Keplerian mass $M=0$, its collapse might just mean its rapid expansion, that
is expansion of the pure scalar field it is made of.

For ingoing matter waves, as before, using Eq.(54) and (75), we get the
index 
\begin{equation}
\overline{N}(r)=\frac{n(r)}{\beta }\sqrt{1-\alpha ^{2}\left( 1+\frac{m^{2}}{%
4r^{2}}\right) ^{2}},
\end{equation}%
which, at the throat\ $r=r_{\text{th}}=\frac{m}{2}$, reduces to%
\begin{equation}
\overline{N}(r_{\text{th}})=\frac{2}{\beta }\sqrt{1-4\alpha ^{2}}.
\end{equation}%
The values of $\alpha <\frac{1}{2}$ and $0<\beta <1$ can be adjusted so that 
$0<\overline{N}(r_{\text{th}})<\infty $. Thus, for $\beta \rightarrow 0$, $%
\alpha <\frac{1}{2}$, it is possible to make $\overline{N}(r_{\text{th}%
})\rightarrow \infty $, so that one has $\overline{R}_{\text{dB}}^{\text{a.o.%
}}\rightarrow 1$, $\overline{T}_{\text{dB}}^{\text{a.o.}}\rightarrow 0$.
This implies that the asymptotic observers will observe stability due to low
asymptotic velocity of ingoing perturbation. For intermediate values of $%
\overline{N}(r_{\text{th}})$, the probabilities will change accordingly, and
the arguments remain similar to the preceding ones. Finally, the near-throat
local observers will perceive the index to be, from Eq.(81),%
\begin{equation}
\text{ }\widehat{N}(r)=\frac{\widetilde{N}(r)}{\beta }=\frac{1}{\beta }\left[
1-\left( \frac{\widetilde{\lambda }}{\lambda _{\text{C}}}\right) ^{2}\right]
^{1/2}\text{,}
\end{equation}%
which implies that $0<$ $\widehat{N}(r)<\infty $. Thus, exactly as above, at
the limit $\beta \rightarrow 0$, it follows that $\widehat{R}_{\text{dB}}^{%
\text{l.o.}}\rightarrow 1$, $\widehat{T}_{\text{dB}}^{\text{l.o.}%
}\rightarrow 0$, and the same arguments just as above apply. Local observers
will observe stability, while asymptotic observers may or may not.

\textbf{8. Phantom wormhole}

Recently, Lobo, Parsaei and Riazi [3] have derived a number of phantom
wormhole solutions. See also [42,43] for a class of earlier solutions, which
have been shown to be stable under perturbations within the class [44]. We
shall here consider only the solution with bounded mass function [3] for
exemplifying the probabilistic coefficients. Throughout the foregoing, we
had used $r$ as the isotropic radial coordinate. The bounded mass function
phantom wormhole they derived is in "standard" coordinates and is given by 
\begin{equation}
ds^{2}=\left( 1+\frac{aR_{0}}{R}\right) ^{1-\frac{1}{a}}c_{0}^{2}dt^{2}-%
\frac{dR^{2}}{1-\frac{R_{0}}{R}\left[ \frac{aR_{0}}{R}+1-a\right] }%
-R^{2}\left( d\theta ^{2}+\sin ^{2}\theta d\varphi ^{2}\right) ,
\end{equation}%
where $-1<a<0$, and $R$ is the standard radial coordinate. The throat
appears at $R_{\text{th}}=R_{0}$. The mass function is%
\begin{equation}
\mathcal{M(}R)=\frac{aR_{0}}{2}\left( \frac{R_{0}}{R}-1\right) .
\end{equation}%
The physical interpretation of the parameter $a$ is that it affects the
redshifts of the signals originating from the throat or its nearby regions
as follows 
\begin{equation}
z=\frac{\delta \lambda }{\lambda }=1-\left( 1+a\right) ^{\frac{1-a}{2a}}%
\text{.}
\end{equation}

We convert the solution to isotropic form by the radial transform $%
R\rightarrow r$ ,%
\begin{equation}
r=\left( \sqrt{R-R_{0}}+\sqrt{R+aR_{0}}\right) ^{2}.
\end{equation}%
The $r-$coordinate radius of the throat is 
\begin{equation}
r_{\text{th}}=(1+a)R_{0}\text{,}
\end{equation}%
and inverting Eq.(98), we find 
\begin{equation}
R(r)=\frac{(1+a)r^{4}-2(a-1)r^{2}r_{\text{th}}+(1+a)r_{\text{th}}^{2}}{%
4(1+a)r^{2}}.
\end{equation}%
The metric (95) can be re-written in the\ isotropic form%
\begin{equation}
ds^{2}=\left[ 1+\frac{aR_{0}}{R(r)}\right] ^{1-\frac{1}{a}%
}c_{0}^{2}dt^{2}-\mu ^{2}[R(r)]\left[ dr^{2}+r^{2}\left( d\theta ^{2}+\sin
^{2}\theta d\varphi ^{2}\right) \right] ,
\end{equation}

\begin{equation}
\mu (R)=\Phi ^{-1}=\frac{R}{\left( \sqrt{R-R_{0}}+\sqrt{R+aR_{0}}\right) ^{2}%
},
\end{equation}%
where $R$ is given by Eq.(100). Therefore, the refractive index, as
perceived by asymptotic observers, is%
\begin{equation}
n[R(r)]=\Phi ^{-1}\Omega ^{-1}=\mu (R)\left[ 1+\frac{aR_{0}}{R}\right] ^{%
\frac{1}{2}\left( \frac{1}{a}-1\right) }
\end{equation}%
and similarly, the index as perceived by near-throat local observers is%
\begin{equation*}
\widetilde{n}[R(r)]=n\Phi =\left[ 1+\frac{aR_{0}}{R}\right] ^{\frac{1}{2}%
\left( \frac{1}{a}-1\right) }
\end{equation*}%
At the throat $R_{\text{th}}=R_{0}$, the indices have the values 
\begin{equation}
n(R_{\text{th}})=(1+a)^{\frac{1}{2a}-\frac{3}{2}},\text{ }\widetilde{n}(R_{%
\text{th}})=(1+a)^{\frac{1}{2a}-\frac{1}{2}}.
\end{equation}%
Thus, as $a\rightarrow -1$, we have $n(R_{\text{th}})$, $\widetilde{n}(R_{%
\text{th}})\rightarrow \infty $, so that $\overline{N}(R_{\text{th}})$, $%
\widehat{N}(R_{\text{th}})\rightarrow \infty $ irrespective of the value of $%
0<\beta <1$, finally leading to 
\begin{eqnarray}
R_{\text{photon}}^{\text{a.o.}} &=&\widetilde{R}_{\text{photon}}^{\text{l.o.}%
}=1\text{, }T_{\text{photon}}^{\text{a.o.}}=\widetilde{T}_{\text{photon}}^{%
\text{l.o.}}=0, \\
\overline{R}_{\text{dB}}^{\text{a.o.}} &=&\widehat{R}_{\text{dB}}^{\text{l.o.%
}}=1\text{, }\overline{T}_{\text{dB}}^{\text{a.o.}}=\widehat{T}_{\text{dB}}^{%
\text{l.o.}}=0.
\end{eqnarray}%
and, as $a\rightarrow 0$, we have $n(R_{\text{th}}),\widetilde{n}(R_{\text{th%
}})\rightarrow \sqrt{e}$, so that likewise 
\begin{eqnarray}
R_{\text{photon}}^{\text{a.o.}} &=&\widetilde{R}_{\text{photon}}^{\text{l.o.}%
}=0.06\text{, }T_{\text{photon}}^{\text{a.o.}}=\widetilde{T}_{\text{photon}%
}^{\text{l.o.}}=0.94, \\
\overline{R}_{\text{dB}}^{\text{a.o.}} &=&\widehat{R}_{\text{dB}}^{\text{l.o.%
}}=0.06\text{, }\overline{T}_{\text{dB}}^{\text{a.o.}}=\widehat{T}_{\text{dB}%
}^{\text{l.o.}}=0.94\text{.}
\end{eqnarray}

It is quite evident from the metric (95) itself that, in the extreme limit $%
a\rightarrow -1$, the throat behaves like the Schwarzschild horizon, where $%
\Omega ^{2}\rightarrow 0$. Consistent with such behavior, the coefficients
in Eqs.(105,106) indicate that, to asymptotic and local observers, the
phantom wormhole will appear almost definitely stable no matter whether the
perturbing pulse is an ingoing photon or a matter particle. On the contrary,
in the other extreme limit, $a\rightarrow 0$, all observers will find the
wormhole to be almost definitely unstable, as the coefficients in
Eqs.(107,108) show. For intermediate values of $a$, the asymptotic and local
observers would see all combinations of coefficients leading to $R+T=1$, so
again there are possibilities that the asymptotic observer will see ghost
phantom wormholes.

The indeterminacy \textit{a la} Tangherlini leading to wormhole
(in)stability might seem somewhat counter-intuitive at first sight, but
would make sense if such indeterminacy could be translated into observable
deterministic signatures of ghost wormholes. However, in the absence of
known experiments on ghost wormholes, we propose below a possible thought
experiment, where we explicitly calculate those signatures.

\textbf{9. A thought experiment}

The experiment exploits, for instance, different gravitational lensing
properties of black and wormholes (for some earlier works on such
properties, see [45-48]). For simplicity, we assume that there is only one
object in the Universe and that local ($\widetilde{O}_{\text{l.o}}$) and
asymptotic observers ($O_{\text{a.o}.}$) exchange information by means of
some signal about their lensing measurements of the object. Further, we
assume that they know the works by Gonz\'{a}lez \textit{et al.} [5,6] about
Ellis wormhole's deterministic collapse into Schwarzschild black hole
corresponding to ($R$, $T$) $=$ ($0$, $1$). On the other hand, as shown
above, the coefficients need not always be exact certainties like ($R$, $T$) 
$=$ ($0$, $1$) or ($1,0$). Hence, both observers can have fractional
probabilities, however big or small, of observing (in)stabilities of the
throat leading to black or wormholes, as the case may be. Both objects can
act as lens ($L$) giving images of background sources ($S$) that are
observed by observers ($O$). When $L,S$ and $O$ are aligned along a line,
the images form a ring called the Einstein ring centered at the lens;
otherwise there will be discrete images on both sides of the optic axis $OL$
at angular locations denoted by $\theta $.

Coming to lensing observations, an essential constraint is required to be
fulfilled: The light coming from the background source to the observer
should not get wound up right on the photon sphere centered at the lens,
with the sphere radius $r=r_{\text{ps}}$ defined by the largest root of the
equation 
\begin{equation}
\frac{1}{A}\frac{dA}{dr}=\frac{1}{C}\frac{dC}{dr},
\end{equation}%
for a generic metric 
\begin{equation}
d\tau ^{2}=A(r)dt^{2}-B(r)dr^{2}-C(r)\left( d\theta ^{2}+\sin ^{2}\theta
d\varphi ^{2}\right) .
\end{equation}%
If the incoming light is eternally captured by the photon sphere, no image
will be formed.

However, the incoming light rays that pass very near to the photon sphere
yield \textit{strong} field lensing observables [45] measured by both the
observers ($\widetilde{O}_{\text{l.o}}$) and ($O_{\text{a.o}.}$). The
easiest observable to evaluate is the minimum impact parameter $u_{m}$, and
this information alone can already distinguish between Schwarzschild and
other types of geometry such as wormholes. Therefore, we consider situations
that guarantee $u_{m}>r_{\text{ps}}$. The parameter $u_{m}$ is proportional
to the separation $\theta _{\infty }$ between each set of relativistic
images with respect to the central lens given by [45]%
\begin{equation}
u_{m}=D_{\text{OL}}\theta _{\infty }\text{,}
\end{equation}%
where $D_{\text{OL}}$ is the distance between the observer ($O$) and the
lens ($L$) and 
\begin{equation}
u_{m}(r_{\text{ps}})=\sqrt{C(r_{\text{ps}})/A(r_{\text{ps}})}.
\end{equation}%
The radius of the photon sphere for the Ellis wormhole metric (52-55) is 
\begin{equation}
r_{\text{ps}}=M\left( 1+\frac{1}{2}\sqrt{4+\frac{1}{\gamma ^{2}}}\right) .
\end{equation}%
Evidently, it is the evaluation of $u_{m}$, or equivalently $\theta _{\infty
}$, defined by the metric functions $C(r)$, $A(r)$ that distinguishes one
spacetime from the other. 

Note that, for fixed $M$, it is only the value of $\gamma $ that determines
the refractive indices ($n$, $\widetilde{n}$), hence the probabilistic
coefficients ($R$,$T$) or ($\widetilde{R}$,$\widetilde{T}$) at the throat,
observed by local and asymptotic observers. These probabilistic coefficients
in turn dictate which observer is likely to observe what $\theta _{\infty }$%
. Thus, all deterministic observations boil down to measuring the values of $%
\gamma $, \textit{as it were}, by truly measuring $\theta _{\infty }$. In
Sec.7(ii), the dependence on $\gamma $ was already demonstrated but the
situation was idealized ($M=1$) without reference to any practical (lensing)
measurement. 

We now consider a practical situation: It is widely believed that the center
of our galaxy hosts a supermassive black hole (BH) with mass $M=2.8\times
10^{6}M_{\odot }$ [45-50]. Taking $D_{\text{OL}}=8.5$ kpc $=2.62\times
10^{22}$ cm as the distance between the local observer ($\widetilde{O}_{%
\text{l.o}}$) identified at the sun and the lens ($L$) at the center of our
galaxy, Virbhadra and Ellis [51] found that $\theta _{\infty }^{\text{BH}%
}\sim 17$ microarcsecs. In principle, such a resolution at the microarcsec
level is reachable by actual VLBI projects in the near future [45].

Now we recall the calculations of Sec.5, where we showed that only the 
\textit{exclusive} value $\gamma =-i$ yielded a Schwarzschild black hole.
Using this imaginary $\gamma $, the values $r_{\text{ps}}=7.73\times 10^{11}$
cm, $u_{m}=2.15\times 10^{12}$ cm and  $\theta _{\infty }^{\text{BH}}=16.943$
microarcsecs can be accurately calculated from Eqs. (111-113), once the
independently observed values of $M$ and $D_{\text{OL}}$ are put in. We also
notice that the condition $u_{m}>r_{\text{ps}}$ is fulfilled. If
measurements confirm this value of $\theta _{\infty }$ within experimental
errors, then $\widetilde{O}_{\text{l.o}}$ concludes a Schwarzschild black
hole. The precision must be very high in order to arrive at this conclusion.
Even a small deviation from this value of  $\theta _{\infty }^{\text{BH}%
}=16.943$ microarcsecs would mean that we are perhaps looking at the
positive mass mouth $M$ of the Ellis wormhole at the galactic center \textit{%
instead} of a black hole. For instance, the measurements of $\widetilde{O}_{%
\text{l.o}}$ might correspond to a real $\gamma $ (wormhole), say, $\gamma
=50$. With the other observed data $M=2.8\times 10^{6}M_{\odot }$ and $D_{%
\text{OL}}=8.5$ kpc remaining the same, he would then measure\ the wormhole
(WH) value $\theta _{\infty }^{\text{WH}}=17.736$ microarcsecs, not too
different from the black hole value $\theta _{\infty }^{\text{BH}}$. The
VLBI missions should have a very high precision to be able to distinguish
such small differences ($\leq 0.793$ microarcsec). Note that $\theta
_{\infty }^{\text{WH}}$ saturates to the value above as $\gamma \rightarrow
\infty $ and so cannot be reduced further. 

To find the observations of the asymptotic observer ($O_{\text{a.o}}$), we
really do not go to $r=\infty $, but just consider him to be located far
away from ($\widetilde{O}_{\text{l.o}}$), e.g., $D_{\text{OL}}=\kappa r_{%
\text{ps}}>$ $8.5$ kpc, where $\kappa =3\times 10^{10}$ (say). Assuming the
scalar field contribution to be small, say $\gamma =0.1$ (wormhole) at this
large distance from the lens, we find the following values: The photon
sphere of the wormhole has a radius $r_{\text{ps}}=2.52\times 10^{12}$ cm,
observer-lens distance $D_{\text{OL}}=\kappa r_{\text{ps}}=24.57$ kpc $%
=7.58\times 10^{22}$ cm and $u_{m}=5.56\times 10^{12}$ cm, which are of the
same order of magnitude as, but about three times larger than, the
Schwarzschild values. Also the condition $u_{m}>r_{\text{ps}}$ is fulfilled.
With the same value $M=2.8\times 10^{6}M_{\odot }$, Eq.(111) now yields  $%
\theta _{\infty }^{\text{WH}}=15.135$ microarcsecs, which is slightly lower
than that of the Schwarzschild value. Measuring this value of $\theta
_{\infty }^{\text{WH}}$ (for that matter, any value differing from the
reference Schwarzschild value  $\theta _{\infty }^{\text{BH}}=16.943$
microarcsecs), $O_{\text{a.o}}$ will conclude a wormhole (since $\gamma $ is
now real). By receiving the Schwarzschild black hole data from $\widetilde{O}%
_{\text{l.o}}$, the distant observer $O_{\text{a.o}}$ further concludes that
the wormhole he sees must be a ghost since, by assumption, there is only one
object and that both observers know that Schwarzschild black hole is only a
result of Ellis wormhole's deterministic collapse [5,6,7].

The observer $O_{\text{a.o}}$ also has the possibility to observe \textit{his%
} black hole value of $\theta _{\infty }$, for which $\gamma =-i$. Then, for
the same mass $M=2.8\times 10^{6}M_{\odot }$ but distance $D_{\text{OL}%
}=24.57$ kpc, he should measure $\theta _{\infty }^{\text{BH}}=5.859$
microarcsecs. Thus, if $O_{\text{a.o}}$ has to observe a black hole, the
accuracy of measurement has to be extremely high, unreachable in the near
future, but cannot be ruled out in principle. All the above arguments show
that the converse can also happen, that is $\widetilde{O}_{\text{l.o}}$ can
observe a wormhole value $\theta _{\infty }^{\text{WH}}=17.736$
microarcsecs, as already discussed above, while $O_{\text{a.o}}$ can observe
a black hole measuring $\theta _{\infty }^{\text{BH}}=5.859$ microarcsecs.
One could then say by reversing the previous arguments that it is $%
\widetilde{O}_{\text{l.o}}$ that sees a ghost wormhole. No ghost phenomenon
appears, when both the observers together measure either real or imaginary
value of $\gamma $, that is, when their conclusions are in agreement.
However, there is no possibility of ghost black holes as further collapse of
a black hole into anything else is not known.

\textbf{10. Conclusions}

It has been concluded in Refs.[5,6,7] on the grounds of metric gravity
master equation that the Ellis wormhole is unstable to linear and non-linear
perturbations. While their deterministic conclusion is correct, we have
revisited in this paper the issue of stability from the viewpoint of local
and asympotic observers applying the results of Tangherlini's
non-deterministic statistical framework dealing with photon motion in real
optical medium. We have extended the application to include also the motion
of matter particles. Tangherlini's approach is essentially heuristic with a
pre-quantum flavor but, interestingly, the expressions of coefficients
[Eqs.(6)] exactly resemble the probabilities obtained by solving the Schr%
\"{o}dinger equation involving a certain potential [36].

It is indeed remarkable that the optics-type equations, such as $p^{\prime
}=n(r)p$ or $p^{\prime }\lambda ^{\prime }=p\lambda =$ constant, are already
non-trivially embedded in the formalism of general relativity expressed in
the language of the medium analogue. Though the analogue is secondary to
metric gravity, it is nonetheless quite useful in that it not only describes
the known general relativistic kinematics in a familiar optical language,
but also leads to new insights in gravity coming from experience in the true
optical regime [10-18]. This success leads us to expect that the application
of Tangherlini's formulation to effective medium could yield genuine
predictions of reflection and transmission probabilities, and by
implication, about wormhole (in)stability back in the general relativity
regime.

The following are our new results:

(i) The expansion of the metric coefficients or the refractive index of the
Ellis wormhole [see Eqs.(61),(62)] show that they can never trivially reduce
to a Schwarzschild black hole for any \textit{real }$\gamma $, unlike in the
case of Ellis I solution (naked singularity). Consequently, Ellis wormhole
would never seem to collapse to a Schwarzschild black hole, contrary to the
"strong" indication stated in Ref.[6]. We show that this need not be the
case. By Wick-like complex transformations, it is possible to show that the
Ellis wormhole can indeed decay into a Schwarzschild black hole and that the
throat would reduce to the horizon.

(ii) We show that the perception of stability of the Ellis wormhole depends
on the location of the observer. When the pulse is an ingoing photon
(Fig.2), to asymptotic observers, the coefficients appear to be $R_{\text{%
photon}}^{\text{a.o.}}\simeq $ $0.60$ and $T_{\text{photon}}^{\text{a.o.}%
}\simeq 0.40$ [Eqs.(64,65)] independently of the size of the throat. This
means\ that those observers\ are more likely to see the ingoing photon
reflected back from the throat providing chances of stability of the Ellis
wormhole. On the other hand, near-throat\ local observers observe the index
as $\widetilde{n}$, which yields the coefficients $\widetilde{R}_{\text{%
photon}}^{\text{l.o.}}\simeq $ $0.20$ and $\widetilde{T}_{\text{photon}}^{%
\text{l.o.}}\simeq 0.80$, meaning that the local observers are most likely
to observe transmission through the throat, hence instability of the
wormhole. The two observations put together give rise to a situation leading
to a ghost wormhole, very similar to a ghost star [Sec.7(ii)]. For massless
Ellis wormhole, the coefficients have the set of values ($R_{\text{photon}}^{%
\text{a.o.}}=\frac{1}{9}$, $T_{\text{photon}}^{\text{a.o.}}=\frac{8}{9}$)
and ($\widetilde{R}_{\text{photon}}^{\text{l.o.}}=0$, $\widetilde{T}_{\text{%
photon}}^{\text{l.o.}}=1$) at the throat. Hence, asymptotic observers cannot
definitively conclude instability, whereas local observers can and there
again emerges the possibility of a zero mass ghost wormhole.

(iii) We have extended the formalism to include de Broglie matter waves such
that $p^{\prime }/p=\overline{N}$ and plotted the corresponding
coefficients. It was shown that the coefficients in general depend not only
on the location of the observer but also on the asymptotic velocity of the
ingoing matter particle. However, for low velocity particles, meaning $\beta
\rightarrow 0$, it follows that $\overline{R}_{\text{dB}}^{\text{a.o.}},%
\widehat{R}_{\text{dB}}^{\text{l.o.}}\rightarrow 1$, implying stability. As
such, for a low momentum impact to the throat, this stability is no
surprise. Fig.3 displays the variation of coefficients with the variation of 
$\beta $. It is seen that, even for $\beta =0.2$, which is not too small a
velocity, the coefficient $\overline{R}_{\text{dB}}^{\text{a.o.}}$could
reach as high a value as $0.76$, [see Sec.7(ii)]. The behavior of $\widehat{R%
}_{\text{dB}}^{\text{l.o.}}$ is exhibited in Fig.4 for some values\ of $%
\beta $. It is evident that local observers are also most likely to see the
wormhole stable for moderate impact at the throat:\ The lower the $\beta $,
the higher is the reflection probability and conversely.

(iv) Wormholes are supposed to have been born in the early universe due to
high energetic processes and ought to have been inflated into macroscopic
size today due to cosmic expansion driven by the phantom energy [52]. The
phantom wormholes recently obtained by Lobo, Parsaei and Riazi [3] could
thus be a real possibility, at least as real as the phantom energy itself.
The question is: Did they survive the bombardment by photons and particles?
Our analysis indicates that extreme phantom wormholes ($a\rightarrow -1$)
should have survived since the coefficients in Eqs.(105,106) indicate that,
to asymptotic and local observers, $R\rightarrow 1$ and the wormhole would
appear almost definitely stable independently of whether the perturbing
pulse is an ingoing photon or a matter particle (Fig.5). For intermediate
values, $-1<a<0$, the asymptotic and local observers would see all possible
combinations of coefficients leading to $R+T=1$, so again there are
possibilities that the asymptotic observer will see ghost phantom wormholes.

The main conclusion is that, as long as the coefficients do not approach
strict values $R=1$, $T=0$, as happens in the case of Schwarzschild black
hole or the extreme phantom wormhole, there is always the possibility that,
while near-throat local observers see instability (low probability of
reflection), the asymptotic observers see stability (high probability of
reflection), thus leading to \textit{ghost wormholes} due to exchange of
observational data between them, as explained in Sec.7(ii).

We wish to comment on some of the challenges: Though the ghost field ($%
\varepsilon =-1$) or phantom energy have come to stay with us for various
reasons, the exotic matter they bring along poses lots of challenges. One is
of course the proven classical instability [5-8,53] threatening the very
survival of wormholes up to today. Mixed configurations such as neutron
star-plus-wormhole systems [54] are also shown to be unstable to linear
perturbations [55]. Exotic matter leads to serious problems at the quantum
level too, where the negative kinetic term leads to the possibility that the
energy density could become arbitrarily negative for high frequency
oscillations [56]. However, the dilatonic Einstein-Gauss-Bonnet wormholes
[57,58] \textit{not} threaded by exotic matter are shown to be stable.
Coming to present work, Tangherlini's non-deterministic arguments seem to
provide a different route to approach the question of stability of wormholes.

A major challenge is posed by a very pertinent question: How to understand
the deterministic validity of the counter-intuitive non-deterministic
results on the (in)stability of Ellis or phantom wormholes?\footnote{%
We thank an anonymous referee for raising this critical question.} This
question arises mainly because the effective medium description is only an
artificial analogue and not a ponderable real medium, so it's not clear how
genuine are the results derived from such an analogue. On the other hand,
such non-deterministic results must have deterministic counterpart for them
to be observationally meaningful. To that end, we discussed in Sec.9, as
unequivocally and convincingly as possible, exactly what numerical values
are to be expected as deterministic counterparts of the indeterminacy under
consideration. Therefore, we argue that the non-deterministic coefficients ($%
R,T$) lurking in the analogue medium is at play throwing up genuine
observable results on the (in)stability of wormholes in the gravity field.

A few cautionary notes seem to be in order: The non-deterministic results
are not \textit{fait accompli} but translate into observable new predictions
yet to be tested by actual experiments. Next, the time scale associated with
this instability is macroscopic as Tangherlini's approach is non-quantum,
the Planck constant $\hslash $ cancelled out of the formulation [see (T3) of
Sec.2]. The scale is of the order of the areal radius of the throat divided
by the speed of light, which is of the order of a few microseconds for a
throat of radius of the order of kilometers. Such time scales have to be
factored in, while devising measurements. Then, we showed in Sec.4 that the
Pound-Rebka frequency shift experiment in the Schwarzschild gravity field is
nothing but the momentum increase of photon in the analogue medium. But this
is purely a heuristic example, not to be taken literally. Since the analogue
medium cannot be seen or touched like the real one, it is only our modest
hope, and not a claim, that different values of observables as an effect of
background indeterminism should nonetheless be measured in the gravity field 
$-$ unless decisively ruled out by actual experiments.

Such a hope is somewhat loosely grounded to a philosophical viewpoint: One
could argue that the whole edifice of theoretical physics is built on
mathematical models that work "unreasonably" well [59]. For example,
Einstein's theory itself is an effective geometric model of gravity $-$
abstract geometry that cannot be seen or touched but its observable
predictions have been accurately tested giving the model the credibility it
deserves. Similarly, our effective medium approach, though a secondary
construct based on \textit{a priori} knowledge of the Ellis wormhole
geometry, would be physically relevant only if the predicted indeterminacy
could be translated into observable deterministic signatures of ghost
wormholes.

Since there are as yet no readymade experiments supporting or refuting such
ghost phenomenon, we devised a possible thought experiment in Sec.9 and
calculated the relevant lensing observable as an attempt to answer the
question posed above. We provided arguments for the deterministic validity
of (in)stability of Ellis/phantom wormholes, the observable signatures of
(in)stability depending on the location of the observers and values of $%
\gamma $. Such location dependent observation is a hallmark of general
relativity and is preserved also in our analysis: It was shown that the
simplest strong lensing observable, the minimum impact parameter $u_{m}$ (or
equivalently $\theta _{\infty }$), measured by observers stationed at
different locations would be different. Thus, while a local observer
measuring $\theta _{\infty }^{\text{BH}}=16.943$ microarcsecs concludes a
Schwarzschild black hole ($\gamma =-i$ strictly), another observer at a
distant location measuring a different value of $\theta _{\infty }^{\text{WH}%
}$ will conclude a wormhole ($\gamma $ real and arbitrary). By exchanging
information between them, the latter observer can deterministically conclude
that he is observing a ghost wormhole. The converse situation is also
possible as explained in Sec.9.

Finally, we remark that the whole preceding development is a consequence of
Tangherlini's ingenious idea of pre-quantum non-deterministic reflection and
transmission coefficients, which are remarkably the same as those obtained
from the quantum Schr\"{o}dinger equation with an appropriate potential
[36]. When a full consistent theory of quantum gravity will be available in
the future, we hope that the concept of ghost wormhole might just manage to
survive.

\textbf{Acknowledgments}

The authors are indebted to Guzel Kutdusova and Arunava Bhadra for useful
conversations. Part of the paper was delivered by one of us (KKN) as an
invited \textit{S.N. Bose Memorial Lecture} on 07 Sept.2015 at the Calcutta
Mathematical Society, India.

\textbf{References}

[1] F.R. Tangherlini, Phys. Rev. A \textbf{12}, 139 (1975).

[2] H.G. Ellis, J. Math. Phys. \textbf{14}, 104 (1973); \textit{Errata}: J.
Math. Phys. \textbf{15}, 520 (1974).

[3] F.S. N. Lobo, F. Parsaei and N. Riazi, Phys. Rev. D \textbf{87}, 084030
(2013).

[4] K.A. Bronnikov, Acta Phys. Polon. B \textbf{4}, 251 (1973).

[5] J.A. Gonz\'{a}lez, F.S. Guzm\'{a}n and O. Sarbach, Class. Quant. Grav. 
\textbf{26}, 015010 (2009).

[6] J.A. Gonz\'{a}lez, F.S. Guzm\'{a}n and O. Sarbach, Class. Quant. Grav. 
\textbf{26}, 015011 (2009).

[7] K.A. Bronnikov, J.C. Fabris and A. Zhidenko, Europhys J. C \textbf{71},
1791 (2011).

[8] K. A. Bronnikov, R. A. Konoplya and A. Zhidenko, Phys. Rev. D \textbf{86}%
, 024028 (2012). See also: K.A. Bronnikov and S. Grinyok, Grav. Cosmol. 
\textbf{10}, 237 (2004); K.A. Bronnikov and A.A. Starobinsky, JETP Lett.%
\textbf{\ 85},1 (2007).

[9] C. Armend\'{a}riz-Pic\'{o}n, Phys. Rev. D \textbf{65}, 104010 (2002).

[10] K.K. Nandi and A. Islam, Am. J. Phys. \textbf{63}, 251 (1995).

[11] A.S. Eddington, \textit{Space, Time and Gravitation }(Cambridge U.P.,
Cambridge, 1920, reprinted 1987).

[12] F. de Felice, Gen. Relat. Grav. \textbf{2}, 347 (1971).

[13] J. Evans, K.K. Nandi and A. Islam, Am. J. Phys. \textbf{64}, 1404
(1996).

[14] J. Evans, P.M. Alsing, S. Giorgetti and K.K. Nandi, Am. J. Phys. 
\textbf{69}, 1103 (2001).

[15] P.M. Alsing, J.C. Evans and K.K. Nandi, Gen. Relat. Grav. \textbf{33},
1459 (2001).

[16] K.K. Nandi, Y.Z. Zhang, P.M. Alsing, J.C. Evans and A. Bhadra, Phys. Rev. D \textbf{67}, 025002 (2003).

[17] J. Evans, K.K. Nandi and A. Islam, Gen. Relat. Grav. \textbf{28}, 413
(1996).

[18] P.M. Alsing, Am. J. Phys.\textbf{\ 66}, 779 (1998).

[19] J. Javanainen and J. Ruostekoski, Phys. Rev. A \textbf{52}, 3033 (1995).

[20] L.V. Hau, S.E. Harris, Z. Dutton and C.H. Behroozi, Nature (London) 
\textbf{397}, 594 (1999).

[21] U. Leonhardt and P. Piwnicki, Phys. Rev. A \textbf{60}, 4301 (1999).

[22] U. Leonhardt and P. Piwnicki, Phys. Rev. Lett. \textbf{84}, 822 (2000); 
\textit{ibid}. \textbf{85}, 5253 (2000).

[23] M. Visser, Phys. Rev. Lett. \textbf{85}, 5252 (2000).

[24] U. R. Fischer and M. Visser, Phys. Rev. Lett.\textbf{\ 88}, 110201
(2002).

[25] Dutton,Z./VestergaardHau,L., Phys. Rev. A \textbf{70}, 053831 (2004).

[26] W. G. Unruh, Phys. Rev. Lett. \textbf{46},1351(1981).

[27] W.G. Unruh, Phys. Rev. D \textbf{51}, 2827 (1995).

[28] M. Visser, Class. Quant. Grav. \textbf{15}, 1767 (1998).

[29] M. Visser, Phys. Rev. Lett. \textbf{80}, 3436 (1998).

[30] S. Liberati, S. Sonego and M. Visser, Class. Quant. Grav. \textbf{17},
2903 (2000).

[31] M. Visser, C. Barcelo and S. Liberati, Gen. Rel. Grav. \textbf{34},
1719 (2002).

[32] J. H. Poynting, Philos. Mag. \textbf{9}, 393 (1905).

[33] R. V. Jones, Nature (London), \textbf{167}, 439 (1951).

[34] R. V. Jones and J. C. S. Richards, Proc. Roy. Soc. A \textbf{221}, 480
(1954).

[35] A. Ashkin and J. M. Dziedzic, Phys. Rev. Lett. \textbf{30},139 (1973).

[36] D. Bohm, \textit{Quantum Theory} (Prentice-Hall, Inc., New York, 1951),
p.235

[37] S.K. Bose, \textit{Introduction to General Relativity} (Wiley Eastern,
New Delhi, 1980).

[38] M.A. Scheel, S.L. Shapiro and S.A. Teukolsky, Phys. Rev. D \textbf{51},
4208 (1995); \textit{ibid.} \textbf{51}, 4236 (1995).

[39] W. Rindler, \textit{Essentials of Relativity} (Springer-Verlag, New
York, 1977), p.189

[40] M.S. Morris and K.S. Thorne, Am. J. Phys. \textbf{56}, 395 (1988).

[41] F. Abe, Astrophys. J. \textbf{725}, 787 (2010).

[42] S.V. Sushkov, Phys. Rev. D \textbf{71}, 043520 (2005).

[43] F.S.N. Lobo, Phys. Rev. D \textbf{71}, 084011 (2005).

[44] F.S.N. Lobo, Phys. Rev. D\textbf{\ 71}, 124022 (2005).

[45] V. Bozza, Phys. Rev. D \textbf{66}, 103001 (2002).

[46] K.S. Virbhadra and G.F.R. Ellis, Phys. Rev. D \textbf{62}, 084003
(2000).

[47] C. R. Keeton and A.O. Petters, Phys. Rev. D \textbf{72},104006 (2005); 
\textit{ibid.} D \textbf{73}, 044024 (2006);\textit{\ ibid.} D\textbf{\ 73},
104032 (2006).

[48] K.K. Nandi, Y.Z. Zhang and A.V. Zakharov, Phys. Rev. D \textbf{74},
024020 (2006).

[49] J.B. Hartle, \ \textit{Gravity - An introduction to general relativity}
(Pearson Inc., San Francisco, 2003).

[50] D. Richstone \textit{et al.}, Nature (London), \textbf{395}, A14 (1998).

[51] K.S. Virbhadra and G.F.R. Ellis, Phys. Rev. D \textbf{62}, 084003
(2000).

[52] S.V. Sushkov and S-W. Kim, Gen. Rel. Grav. \textbf{36},1671 (2004).

[53] H. -a. Shinkai and S. A. Hayward, Phys. Rev. D \textbf{66}, 044005
(2002).

[54] V. Dzhunushaliev, V. Folomeev, B. Kleihaus and J. Kunz, Phys. Rev. D 
\textbf{85}, 124028 (2012).

[55] V. Dzhunushaliev, V. Folomeev, B. Kleihaus and J. Kunz, Phys. Rev. D 
\textbf{87}, 104036 (2013).

[56] S. V. Sushkov and Y.-Z. Zhang, Phys. Rev. D \textbf{77}, 024042 (2008).

[57] P. Kanti, B. Kleihaus and J. Kunz, Phys. Rev. Lett. \textbf{107},
271101 (2011).

[58] P. Kanti, B. Kleihaus and J. Kunz, Phys. Rev. D \textbf{85}, 044007
(2012).

[59] M. Steiner, \textit{The applicability of mathematics as a philosophical
problem} (Harvard University Press, Massachusetts, 1998).

\begin{center}
---------------------------------------
\end{center}

\textbf{Figure Captions}

\begin{figure} 
\centerline { \includegraphics 
[scale=0.8]{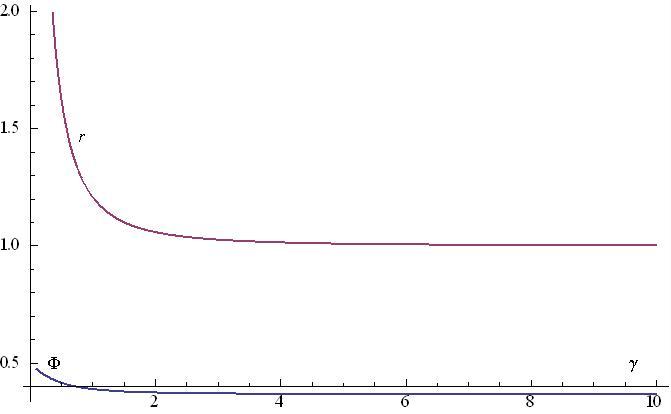}}
\caption{Plot of $\Phi (r_{\text{th}})$ and $r_{\text{th}}(\gamma )$ vs $%
\gamma $. The larger the value of $\gamma $, the smaller the value of $r_{%
\text{th}}$ and $\Phi (r_{\text{th}})$. One should choose $\alpha $ such
that $\alpha <\Phi (r_{\text{th}})$ needed for the reality of $N(r)$.}
\end{figure}

\begin{figure}
\centerline { \includegraphics 
[scale=0.8]{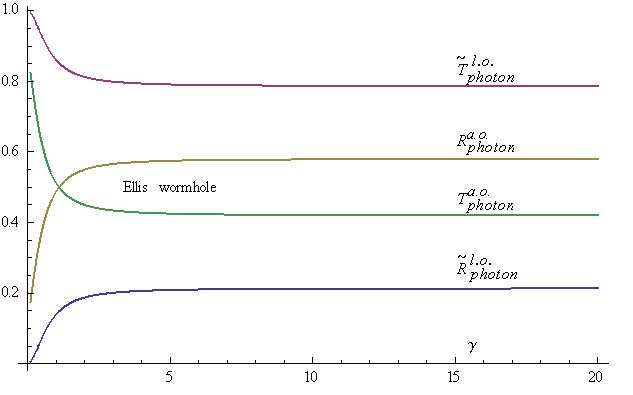}}
\caption{Reflection ($R$) and Transmission ($T$) probabilities for the ingoing
light pulse in the Ellis wormhole spacetime. We choose units such that $M=1$
and plot ($R,T$) at the throat, now only a function of $\gamma $ for
different location of observers [Eqs.(64-67)]. Asymptotic observers are
denoted by the superscript (a.o.) and local observers by (l.o.) with a tilde
for coefficients. It turns out that $R$ and $T$ reach fixed values at some
value of $\gamma $, hence at some fixed size of the throat $r_{\text{th}}$,
where $R_{\text{photon}}^{\text{a.o.}}\simeq 0.60$, $T_{\text{photon}}^{%
\text{a.o.}}\simeq 0.40$, and $\widetilde{R}_{\text{photon}}^{\text{l.o.}%
}\simeq $ $0.20$ and $\widetilde{T}_{\text{photon}}^{\text{l.o.}}\simeq 0.80$%
, as shown in the figure. The values are in contrast to the Schwarzschild
values $R=1$, $T=0$ at the horizon, which suggests that the horizon plays
the role of a perfect mirror totally reflecting away the incoming pulse.}
\end{figure}

\begin{figure}[tbp]
\centerline { \includegraphics 
[scale=0.8]{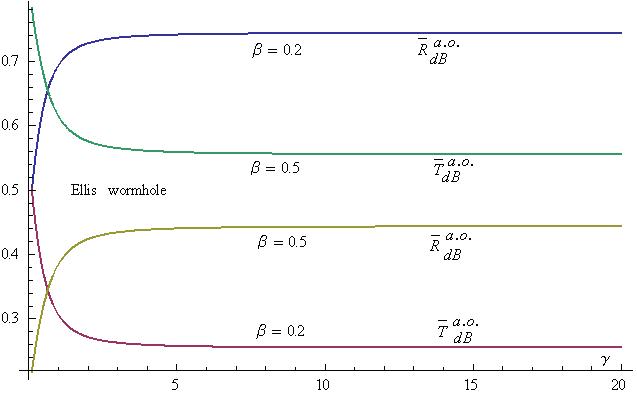}}
\caption{Reflection ($R$) and Transmission ($T$) probabilities for ingoing de
Broglie waves in the Ellis wormhole spacetime as perceived by the asymptotic
observers. We choose units such that $M=1$ and values $\alpha =\lambda
^{\prime }/\lambda _{\text{C}}=0.01$, $\beta =0.2$ and $0.5$. The~plot ($R,T$%
) at the throat, now only a function of $\gamma $ [Eqs.(78,79)], shows that $%
R$ and $T$ \ reach fixed values at some value of $\gamma $, hence at some
fixed size of the throat $r_{\text{th}}$. Asymptotic observers are denoted
by the superscript (a.o.) for whom $\overline{R}_{\text{dB}}^{\text{a.o.}%
}\sim 0.76$, $\overline{T}_{\text{dB}}^{\text{a.o.}}\sim 0.24$, when $\beta
=0.2$. Reflection probability is reduced to $\overline{R}_{\text{dB}}^{\text{%
a.o.}}=0.45$, when $\beta $ is increased, e.g., to $0.5$, as shown in the
figure. When $\beta \rightarrow 0$, stability is achieved: $\overline{R}_{%
\text{dB}}^{\text{a.o.}}\rightarrow 1$, $\overline{T}_{\text{dB}}^{\text{a.o.%
}}\rightarrow 0$.}
\end{figure}

\begin{figure}[tbp]
\centerline { \includegraphics 
[scale=0.8]{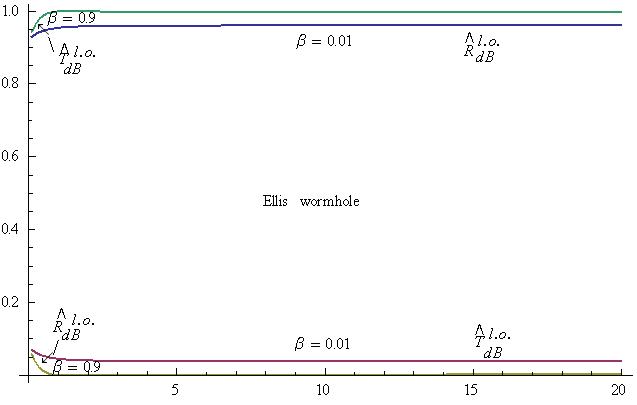}}
\caption{Reflection ($R$) and Transmission ($T$) probabilities for ingoing de
Broglie waves in the Ellis wormhole spacetime perceived by the near-throat
local observers. The~plot ($R,T$) at the throat, now only a function of $%
\gamma $ [Eqs.(83,84)], shows de Broglie waves of high and low velocities $%
\beta =0.9$ and $0.01$. The lesser the velocity, the higher is the $\widehat{%
R}_{\text{dB}}^{\text{l.o.}}$. Stability is achieved: $\widehat{R}_{\text{dB}%
}^{\text{l.o.}}\rightarrow 1$,$\widehat{T}_{\text{dB}}^{\text{l.o.}%
}\rightarrow 0$ only in the limit $\beta \rightarrow 0$.}
\end{figure}

\begin{figure}[tbp]
\centerline { \includegraphics 
[scale=0.8]{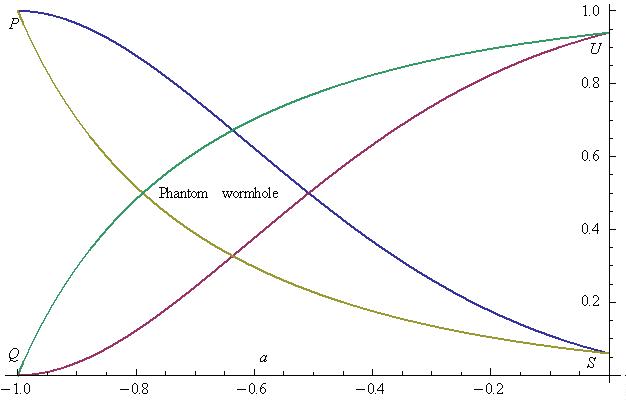}}
\caption{Reflection ($R$) and Transmission ($T$) probabilities for de Broglie
waves in the phantom wormhole spacetime ($-1<a<0$). The point $P$
corresponds to $R_{\text{photon}}^{\text{a.o.}}=\widetilde{R}_{\text{photon}%
}^{\text{l.o.}}=1$, $\overline{R}_{\text{dB}}^{\text{a.o.}}=\widehat{R}_{%
\text{dB}}^{\text{l.o.}}=1$ and $Q$ corresponds to $T_{\text{photon}}^{\text{%
a.o.}}=\widetilde{T}_{\text{photon}}^{\text{l.o.}}=0$, $\overline{T}_{\text{%
dB}}^{\text{a.o.}}=\widehat{T}_{\text{dB}}^{\text{l.o.}}=0$. See
Eqs.(105,106). Likewise, $S$ corresponds to $R_{\text{photon}}^{\text{a.o.}}=%
\widetilde{R}_{\text{photon}}^{\text{l.o.}}=0.06,$ $\overline{R}_{\text{dB}%
}^{\text{a.o.}}=\widehat{R}_{\text{dB}}^{\text{l.o.}}=0.06$ and $U$
corresponds to $T_{\text{photon}}^{\text{a.o.}}=\widetilde{T}_{\text{photon}%
}^{\text{l.o.}}=0.94$, $\overline{T}_{\text{dB}}^{\text{a.o.}}=\widehat{T}_{%
\text{dB}}^{\text{l.o.}}=0.94$. See Eqs.(107,108).}
\end{figure}

\end{document}